\documentclass[referee]{aa} 
\usepackage{graphicx}
\newcommand{\kms}{${\rm km \: s^{-1}}$}
\begin{document}

\title{Oxygen line formation in late-F through early-K disk/halo stars
\thanks{Tables 2, 5, and 7 are available only in electronic form,
which are temporarily at the anonymous ftp site of
{\tt ftp://133.11.160.242/Users/takeda/aams2727/}
but will be replaced (submitted) to CDS when the paper is 
to be published.}
}

\subtitle{Infrared O~{\sc i} triplet and [O~{\sc i}] lines}

\author{Y. Takeda}
\institute{Komazawa University, Setagaya, Tokyo 154-8525, Japan\\
\email{takedayi@cc.nao.ac.jp}
}

\abstract{
In order to investigate the formation of O~{\sc i} 7771--5 and 
[O~{\sc i}] 6300/6363 lines, extensive non-LTE calculations for 
neutral atomic oxygen were carried out for wide ranges of model 
atmosphere parameters, which are applicable to early-K through 
late-F halo/disk stars of various evolutionary stages.

The formation of the triplet O~{\sc i} lines was found to be
well described by the classical two-level-atom scattering model,
and the non-LTE correction is practically determined by the
parameters of the line-transition itself without any significant
relevance to the details of the oxygen atomic model.
This simplifies the problem in the sense that the non-LTE
abundance correction is essentially determined only by 
the line-strength ($W_{\lambda}$), if the atmospheric parameters 
of $T_{\rm eff}$, $\log g$, and $\xi$ are given, without any
explicit dependence of the metallicity; thus allowing a useful
analytical formula with tabulated numerical coefficients.
On the other hand, our calculations lead to the robust conclusion 
that LTE is totally valid for the forbidden [O~{\sc i}] lines.

An extensive reanalysis of published equivalent-width data of 
O~{\sc i} 7771--5 and [O~{\sc i}] 6300/6363 taken from various 
literature resulted in the conclusion that, while
a reasonable consistency of O~{\sc i} and [O~{\sc i}] abundances was observed 
for disk stars ($-1 \la [{\rm Fe}/{\rm H}] \la 0$), the existence of 
a systematic abundance discrepancy was confirmed between O~{\sc i} and 
[O~{\sc i}] lines in conspicuously metal-poor halo stars
($-3 \la [{\rm Fe}/{\rm H}] \la -1$) without being removed by 
our non-LTE corrections, i.e., the former being larger by $\sim 0.3$ dex 
at $-3 \la [{\rm Fe}/{\rm H}] \la -2$.

An inspection of the parameter-dependence of this discordance
indicates that the extent of the discrepancy tends to be
comparatively lessened for higher $T_{\rm eff}/\log g$ stars,
suggesting the preference of dwarf (or subgiant) stars for 
studying the oxygen abundances of metal-poor stars.


}

\maketitle

\section{Introduction}

Oxygen abundance determination in F--K stars is one of the most 
controversial and interesting topics in stellar spectroscopy,
especially concerning the behavior of the [O/Fe] ratio in metal-poor
halo stars (see, e.g., Carretta et al. 2000; King 2000; Israelian et al.
2001; Nissen et al. 2002; and the references therein). 

Namely, we are faced with a problem currently in hot controversy 
on the observational side, which stems from the fact that each of the 
different groups of spectral lines, 
(a) high-excitation permitted O~{\sc i} lines (e.g., 7773 triplet), 
(b) low-excitation forbidden [O~{\sc i}] lines (6300/6363 lines), and 
(c) OH lines at UV ($\sim$ 3100--3200 $\rm\AA$), 
do not necessarily yield reasonably consistent oxygen abundances 
in metal-poor stars.

More precisely, a distinct discrepancy is occasionally reported between
the abundances derived from O~{\sc i} and [O~{\sc i}] lines in the sense that 
the former tend to yield systematically higher oxygen abundances
than the latter in metal-poor stars (see, e.g., Cavallo et al. 1997),
in contrast to the relation between O~{\sc i} and OH lines which give
more or less consistent abundances (e.g., Boesgaard et al. 1999).
As a result, different trends of O-to-Fe ratio, an important tracer
of early nucleosynthesis in our Galaxy, have naturally been suggested:
(1) plateau-like [O/Fe] from [O~{\sc i}] lines, and (2) ever-rising [O/Fe]
from O~{\sc i} or OH lines.

Admittedly, what has been described above is nothing but a rough summary
and the actual situation is more complicated than such a simple
dichotomic picture. 
For example, the O~{\sc i} vs. [O~{\sc i}]
discordance appears to be observed only in {\it metal-poor} halo stars,
since such a systematic disagreement is absent in population I G--K giants
(cf. Takeda et al. 1998).
Moreover, according to Carretta et al.'s (2000) 
recent investigation of very metal-poor stars (mainly giants), 
a ``nearly flat'' [O/Fe] trend was concluded consistently
from both of the O~{\sc i} and [O~{\sc i}] lines. 
Very recently, on the other hand, Nissen et al. (2002) also
obtained nearly consistent O~{\sc i} and [O~{\sc i}] abundances
of metal-poor dwarfs and subgiants, but at a ``quasi-linearly increasing''
trend in contrast to Carretta et al. (2000), 
though an application of the 3D hydrodynamical model 
atmospheres\footnote{
This effect is another controversial factor of importance in the
oxygen abundance problem, and the results from the classical
1D model atmospheres neglecting this effect 
(e.g., the temperature of the upper layer is considerably lowered
in the 3D atmospheric model) may suffer appreciable corrections.
Though theoretical investigations in this field are still in progress,
the nature of the 3D abundance corrections ($\Delta_{\rm 3D}\log\epsilon$)
to be applied to the 1D
solutions for each oxygen line may be roughly summarized as follows
(e.g., Kiselman \& Nordlund 1995; Asplund 2001; Allende Prieto et al. 2001;
Asplund \&  Garc{\'\i}a Per{\'e}z 2001; Nissen et al. 2002):\\
--- This 3D effect is practically negligible for the deeply-forming
high-excitation O~{\sc i} lines (cf. Fig. 10 in Kiselman \& Nordlund 1995;
Asplund 2001).\\
--- Similarly, its effect on the forbidden [O~{\sc i}] line is generally
minor ($|\Delta_{\rm 3D}\log\epsilon| \la 0.1$ dex; the sign of 
the correction is 
negative), but may be increased to an appreciable amount up to $\sim 0.2$ 
dex in lower gravity/metallicity atmospheres (cf. Table 6 in Nissen et al.
2002).\\
--- The 3D effect is most important on the UV OH lines, and conspicuously
large corrections amounting to $|\Delta_{\rm 3D}\log\epsilon| \sim 0.6$ dex 
or even more (at [Fe/H] = $-3.0$; $\Delta_{\rm 3D}\log\epsilon$ being negative)
are suggested (cf. Table 2 in Asplund \&  Garc{\'\i}a Per{\'e}z 2001).\\
Since our main concern in this paper is to investigate the non-LTE effect
on the abundances derived from the O~{\sc i} and [O~{\sc i}] lines, 
our use of the classical 1D models would not cause any serious problems
(actually, taking account of the 3D correction does not essentially 
affect our final conclusion; cf. the footnote 11 in Sect. 4.5.1).
} 
(taking into account the atmospheric granular motions)
tends to deteriorates the agreement.

This confusion makes us wonder which on earth is the correct 
and reliable indicator of the oxygen abundance.
Generally speaking, a modest view currently preferred by many people is 
that [O~{\sc i}] lines are less problematic and more creditable due to
their weak $T_{\rm eff}$-sensitivity as well as the general belief
of the validity for the assumption of LTE,
while O~{\sc i} and OH lines are less reliable because of the uncertainties
involved with the non-LTE correction (O~{\sc i}) or with the 3D effect of
atmospheric inhomogeneity (OH).

Recently, however, we cast doubt on this conservative picture
and suggested the possibility of erroneously underestimated
abundances from [O~{\sc i}] lines (i.e., the problem may be due to 
the formation of forbidden lines rather than that of permitted lines)
based on our analysis of seven metal-poor stars 
(Takeda et al. 2000), which indicated the tendency of above-mentioned 
O~{\sc i} vs. [O~{\sc i}] discrepancy (though a small number of sample 
stars prevented us from making any convincing argument).

Motivated by that work suggesting the necessity of 
a follow-up study to check if our hypothesis is reasonable or not, 
we decided to revisit this O~{\sc i} vs. [O~{\sc i}] problem while paying 
special attention to the topic of the non-LTE correction to be applied,
one of the controversial factors in oxygen abundance determination.
Toward this aim, we carried out new extensive non-LTE calculations,
based on which the numerous published equivalent-width data were 
reanalyzed to derive the non-LTE abundances, in order to discuss 
whether or not the abundance discrepancy between O~{\sc i} and [O~{\sc i}] 
really exists, which we hope to shed some light on the oxygen problem of 
metal-deficient stars mentioned above.

This paper is organized as follows:\\
--- First, extensive tables of non-LTE corrections for 
the O~{\sc i} 7771--5 triplet lines over a wide range of 
atmospheric parameters are presented, so that they may be applied 
to late-F through early-K dwarfs/giants of various metallicities.\\
--- Second, in order for application of such tables to actual analysis of
observational data, we try to find out a useful analytical
formula for evaluating non-LTE corrections of these O~{\sc i} triplet lines
for any given atmospheric parameters ($T_{\rm eff}$, $\log g$, $\xi$)
and an observed equivalent width ($W_{\lambda}$).\\
--- Third, we investigate the possibility of a non-LTE effect for the
low-excitation forbidden [O~{\sc i}] 6300/6363 lines, for which few such
attempt has ever been made.\\
--- Fourth, equivalent-width data of O~{\sc i} 7771--5 lines as well as
[O~{\sc i}] 6300/6363 lines taken from various literature are reanalyzed
within the framework of the consistent system of analysis along with 
application of our non-LTE correction formula, in order to discuss 
the characteristics of [O/Fe] vs. [Fe/H] relation derived from 
permitted O~{\sc i} lines and forbidden [O~{\sc i}] lines separately.

Also, as a supplementary subject related to this study,
the influence of the treatment of H~{\sc i} collision on the non-LTE
corrections of O~{\sc i} 7771--5 lines was quantitatively estimated,
and the consistency of the present choice of this parameter was checked
by comparing the abundances of nearby solar-type stars derived from
O~{\sc i} 7771--5, O~{\sc i} 6158, and [O~{\sc i}] 6300 lines.
In addition, we also tried to qualitatively explain the line formation 
of O~{\sc i} 7771--5 in terms of the simple classical two-level-atom model, 
which gives us an insight to understanding the behavior of 
the non-LTE correction for these triplet lines.
These are separately described in Appendices A and B, respectively.

\section{Statistical equilibrium calculations}

\subsection{Atomic model}

The procedures of our non-LTE calculations for neutral atomic oxygen
are almost the same as those described in Takeda (1992), which
should be consulted for details. We only mention here that
the calculation of collision rates in rate equations basically
follows the recipe described in Takeda (1991). Especially,
the effect of H~{\sc i} collisions, an important key factor in the non-LTE 
calculation of F--K stars, was treated according to Steenbock \& Holweger's
(1984) classical formula {\it without any correction} as has been done 
in our oxygen-related works so far, which we believe to be a reasonable 
choice as far as our calculations are concerned (cf. Takeda 1995,
and Appendix A).

There is, however, an important change concerning the atomic model 
as described below.
The model oxygen atom adopted in Takeda (1992) was constructed mainly
based on the atomic data (energy level data and $gf$ values) compiled by 
Kurucz \& Peytremann (1972), with $\sim$~30 additional radiative transitions
which were missing in their compilation [e.g., resonance transitions
whose transition probabilities were taken from Morton (1991)].
This old model did not include the
2s$^{2}$~2p$^{4}$ $^{3}$P--2p$^{4}$~$^{1}$D transition 
(corresponding to [O~{\sc i}] 6300/6363 lines) explicitly as 
the radiative transition (i.e., it was treated as if being 
completely radiatively forbidden in spite of its actually 
weak connection), simply because it was not contained in 
Kurucz \& Peytremann (1972). 
We also noticed that several important UV radiative transitions
originating upward from the 2p$^{4}$~$^{1}$D term (i.e., upper term of 
[O~{\sc i}] 6300/6363) were also missing in that old model atom 
for the same reason.

We, therefore, newly reconstructed the model oxygen atom exclusively 
based on the data of Kurucz \& Bell (1995), with additional transition 
probability data for 2p$^{4}$~$^{1}$D--3s~$^{3}$S$^{\rm o}$ and
2p$^{4}$~$^{1}$D--3s~$^{5}$S$^{\rm o}$ transitions taken from 
Vienna Atomic Line Database (VALD; Kupka et al. 1999).
Regarding the collisional rate $C_{1-2}$ for the [O~{\sc i}] transition
(2s$^{2}$~2p$^{4}$ $^{3}$P--2p$^{4}$~$^{1}$D) mentioned above,
the cross-section given in Osterbrock (1974) were used for evaluating the
electron collision rates ($C_{1-2}^{\rm e}$), and the H~{\sc i} collision
rates ($C_{1-2}^{\rm H}$) were estimated by scaling $C_{1-2}^{\rm e}$ 
(cf. Takeda 1991).
The resulting atomic model comprises 87 terms 
(including terms up to 3p$''$~$^{1}$S, 6d$'$~$^{3}$P, 10d~$^{5}$D
for the singlet, triplet, and quintet system, respectively) 
and 277 radiative transitions,
which is similar to the old one in terms of the complexity but must be
significantly improved, especially for studying the formation 
of [O~{\sc i}] lines (i.e., more realistic treatment of the 1--2 interaction
and the inclusion of as many UV transitions connecting to the upper term 2
as possible).

Given the changes explained above, the reader can find the description 
on the adopted rates in the statistical-equilibrium calculation
in Sect. 2.1 of Takeda (1992). Some additional details (which were not
explicitly described therein) are given below:\\
--- The population of neutral hydrogen (necessary for computing the
rates of H~{\sc i} collision or the charge-exchange reaction) was
taken from the model atmospheres (i.e., LTE ionization equilibrium).\\
--- For the upper terms other than the lowest 8 terms mentioned 
in Sect. 2.1 of Takeda (1992), the hydrogenic approximation was used
for the photoionization cross section.\\
--- The photoionizing radiation field was computed based on Kurucz's (1993a)
ATLAS9 model atmospheres while incorporating the line opacity with the help
of Kurucz's (1993b) opacity distribution function.\\
--- The collisional rates for those transitions, which are not explicitly 
mentioned in Sect. 2.1 of Takeda (1992), were evaluated according to 
the procedure described in Sect. 3.1.3 of Takeda (1991).

\subsection{Model atmospheres}

We planned to make our calculations applicable to 
stars from near-solar metallicity (population I) down to very low
metallicity (extreme population II) at late-F through early-K spectral 
types in various evolutionary stages (i.e., dwarfs, subgiants, giants, 
and supergiants). We, therefore, decided to carry out non-LTE calculations
on the extensive grid of one hundred ($5 \times 5 \times 4$) 
model atmospheres resulting from combinations of five $T_{\rm eff}$ values 
(4500, 5000, 5500, 6000, 6500 K), five $\log g$ (cm~s$^{-2}$) values 
(1.0, 2.0, 3.0, 4.0, 5.0), and four metallicities (represented by [Fe/H])
(0.0, $-1.0$, $-2.0$, $-3.0$).
As for the stellar model atmospheres, we adopted Kurucz's (1993a) ATLAS9 
models\footnote{It has been occasionally argued that the new treatment
of convection in these ATLAS9 models may not be adequate (e.g., 
Castelli et al. 1997) in the sense that the newly incorporated effect 
of convective overshooting had better be rather switched off 
[like the old Kurucz's (1979) ATLAS6 models]. 
In order to investigate this problem, test calculations were carried out
by using the models without convective overshooting (generated by the
ATLAS9 program while setting OVERWT = 0) and the abundance variations 
between the overshooting-on and -off cases were examined for a wide range 
of atmospheric parameters (5000~K $\leq T_{\rm eff} \leq$~6500K, 
$3 \leq \log g \leq 5$, and $-3 \leq [{\rm Fe}/{\rm H}] \leq 0$).
It turned out that the differences of $\log\epsilon$(ATLAS9 model with
overshooting) $-$ $\log\epsilon$(no-overshooting model) are mostly
positive for both cases of O~{\sc i} and [O~{\sc i}], and quantitatively 
insignificant; i.e., $\la 0.05$ dex for [O~{\sc i}] and $\la 0.1$ dex 
for O~{\sc i} (which are interpreted as reflecting the change 
in the temperature gradient). Considering the same sign of these 
variations, we may consider that this effect can be neglected 
in the present O~{\sc i} vs. [O~{\sc i}] problem.}
corresponding to the microturbulent velocity ($\xi$) of 2 \kms .

\subsection{Oxygen abundance and microturbulence}

Regarding the oxygen abundance used as an input value in non-LTE calculations, 
we assumed $\log\epsilon_{\rm O}^{\rm input}$ = 8.93 + [Fe/H] +[O/Fe],
where [O/Fe] = 0.0 for the solar metallicity models ([Fe/H] = 0) 
and [O/Fe] = +0.5 for the other metal-deficient models 
([Fe/H] = $-1$, $-2$, and $-3$). Namely, the solar oxygen abundance 
of 8.93 taken from Anders \& Grevesse (1989), which is also the canonical
value used in the ATLAS9 models, was adopted for the normal-metal models, 
while the metallicity-scaled oxygen abundance 
plus 0.5 dex (allowing for the characteristics of $\alpha$ element)
was assigned to the metal-poor models.

The microturbulent velocity (appearing in the line-opacity calculations
along with the abundance) was assumed to be 2 \kms , to make it 
consistent with the model atmosphere.

\section{Non-LTE corrections}

\subsection{Evaluation procedures}

Here, we describe how the non-LTE abundance corrections are evaluated
after the departure coefficients, $b \equiv n_{\rm NLTE}/ n_{\rm LTE}^{*}$,
have been established by statistical equilibrium calculations.\\

For an appropriately assigned oxygen abundance ($A^{\rm a}$) and
microturbulence ($\xi^{\rm a}$), we first 
calculated the non-LTE equivalent width ($W^{\rm NLTE}$) of the line 
by using the computed non-LTE departure coefficients ($b$) for each model 
atmosphere. Next, the LTE ($A^{\rm L}$) and NLTE ($A^{\rm N}$) abundances
were computed from this $W^{\rm NLTE}$
while regarding it as if being a given observed equivalent width.
We can then obtain the non-LTE abundance correction, $\Delta$, which is 
defined in terms of these two abundances as 
$\Delta \equiv A^{\rm N} - A^{\rm L}$.

Strictly speaking, the departure coefficients [$b(\tau)$] for 
a model atmosphere correspond to the oxygen abundance and 
the microturbulence of $\log\epsilon_{\rm O}^{\rm input}$
and 2 \kms\ adopted in the calculations (cf. Sect. 2.3). 
Nevertheless, considering the fact that the departure coefficients
(i.e., {\it ratios} of NLTE to LTE number populations) are
(unlike the population itself) not much sensitive to small changes in 
atmospheric parameters, we applied such computed $b$ values also to 
evaluating $\Delta$ for slightly different $A^{\rm a}$ and $\xi^{\rm a}$ 
from those fiducial values assumed in the statistical equilibrium 
calculations.\footnote{The error caused by this practical treatment
was checked based on exact calculations for the representative case of
$T_{\rm eff}$ = 6000~K, $\log g =$ 3.0, and [O/Fe] = [Fe/H] = 0.0,
where the non-LTE correction is rather large ($\Delta \sim -0.5$ dex).
It was confirmed that the $|$correct treatment $-$ practical treatment$|$ 
difference in evaluating the non-LTE correction is quantitatively 
insignificant; i.e., $\la 0.01$ dex 
[the use of $b$ (2 \kms ) for computing $\Delta (2\pm 1)$ \kms ]
and $\la 0.04$ dex [the use of $b$($\log\epsilon_{\rm O}^{\rm input}$)
for computing $\Delta$($\log\epsilon_{\rm O}^{\rm input} \pm 0.3$)].
}
Hence, we evaluated $\Delta$ for three $A^{\rm a}$ values
($\log\epsilon_{\rm O}^{\rm input}$ and $\pm 0.3$ dex perturbation)
as well as three $\xi$ values (2 \kms\ and $\pm 1$ \kms perturbation)
for a model atmosphere using the same departure coefficients.

\subsection{Results of computed corrections}

The computation of $\Delta$ values was carried out for five lines:
three O~{\sc i} lines of 7771--5 triplet and two [O~{\sc i}] lines 
of 6300/6363 doublet.
The WIDTH9 program, which was originally written by R. L. Kurucz 
but modified by the author in various respects (e.g., incorporation
of the non-LTE departure coefficients into the line opacity
and the line source function), was used for calculating 
the equivalent width for a given abundance, or inversely evaluating 
the abundance for an assigned equivalent width.
The adopted line data are summarized in Table 1.

The full results of the computed  $\Delta$ values ($\xi$ = 1, 2, and 3 \kms\ 
cases for each of the three triplet lines at 7771.94, 7774.17, and 
7775.39~$\rm\AA$) are given in Table 2.\footnote{Available only in 
electronic form (cf. the footnote to the title). Similarly, the 
non-LTE corrections for another line, O~{\sc i} 6158, were calculated
for the [Fe/H] = 0 and $-1$ models for the discussion in 
Appendix A, and are presented as (electronic) Table 7.}

As will be described in Sect. 4.3, the non-LTE corrections for the 
forbidden [O~{\sc i}] 6300/6363 lines turned out to be essentially 
equal to zero, suggesting the firm validity of LTE.

\subsection{Analytical formula for practical application}

It is necessary for us to evaluate the non-LTE corrections corresponding to
actual equivalent-width data for stars having a wide variety of 
atmospheric parameters ($T_{\rm eff}$, $\log g$, $\xi$) and metallicity 
([Fe/H]), in order to compare them with those derived by others, or
to apply them to reanalysis of published data.
However, since the raw tabulated values given for each of the individual 
models (e.g., Table 2) are not necessarily convenient for such practical 
applications, it is desirable to develop a useful formula for this purpose.

This motivation reminded us of the simple unique correlation between 
$W_{\lambda}$(O~{\sc i} 7774.17) and its NLTE correction in metal-poor 
solar-type dwarfs (5500~K $\leq T_{\rm eff} \leq$~6500K, 
$3.5 \leq \log g \leq 4.5$, and $-2 \leq [{\rm Fe}/{\rm H}] \leq 0$)
discussed in Takeda (1994; cf. Fig. 10 therein).\footnote{This tendency 
can be interpreted as being due to the close connection between 
the extent of the dilution of the line source function
(determining the extent of the non-LTE effect) and the strength of
these triplet lines (cf. Sect. 4.2 in Takeda 1994, or Appendix B).}
If such a relation really holds also in the present case irrespective 
of the atmospheric parameters (much wider range than that in Takeda 1994), 
it would greatly simplify our task.
However, according to the $\Delta$ vs. $W^{\rm NLTE}$ relation for 
O~{\sc i} 7774.17 depicted in Fig. 1 (based on the data in Table 2),
the situation is not such simple; i.e., $\Delta$ is not only a function
of $W^{\rm NLTE}$ but also more or less dependent\footnote{This dependence 
on $T_{\rm eff}$ and $\log g$ may be qualitatively explained by 
the simple two-level-atom model, as shown in Appendix B.} on $T_{\rm eff}$ 
and $\log g$ (and $\xi$).
Nevertheless, a close examination revealed that for given 
$T_{\rm eff}$, $\log g$, and $\xi$, the extent of the non-LTE correction 
($\Delta$) is a nearly monotonic function of the equivalent width 
($W_{\lambda}$; measured in m$\rm\AA$), and that
\begin{equation}
\Delta = a  10^{b W_{\lambda}}, 
\end{equation}
where $a$ and $b$ are appropriately chosen coefficients 
as functions of ($T_{\rm eff}$, $\log g$, and $\xi$),\\
is a fairly good approximation for an appropriately chosen set of $(a, b)$,
as far as the line is not too strong (i.e., 
$W_{\lambda} \la 100 {\rm m\AA}$).
It should be noted that this relation does not {\it explicitly}
contain the metallicity, though stars with less metals generally show 
smaller extent of $|\Delta|$ values due to their weaker line-strengths 
($W_{\lambda}$).

We therefore determined the best-fit coefficients $(a, b)$ 
based on the computed $\Delta$ values in Table 2 for 
each combination of ($T_{\rm eff}$, $\log g$, $\xi$);
the resulting coefficients are summarized in Tables 3 for each of 
the O~{\sc i} triplet lines at 7771.94, 7774.17, and 7775.39 $\rm\AA$, 
respectively. The analytical approximations presented by Eq. (1) 
with the coefficients given in Table 3 are also depicted 
in Fig. 1 (the solid lines).

Practically, we first evaluate ($a, b$) by interpolating this table 
in terms of $T_{\rm eff}$, $\log g$ and $\xi$ of a star in consideration,
which are then applied to Eq. (1) with the observed equivalent 
width to obtain the non-LTE correction.

In order to demonstrate the applicability of this formula,
we show the error of Eq. (1) (defined as 
$\delta_{\rm error} \equiv \Delta^{\rm formula} - \Delta^{\rm actual}$) 
as a function of $W_{\lambda}$ for the case of O~{\sc i} 7774.17 in Fig. 2
(compare it with Fig. 1), where we can see that an accuracy within 
a few hundredths dex is accomplished as far as 
$W_{\lambda} \la 100$~m$\rm\AA$.

\section{Discussion}

\subsection{NLTE correction for O~I triplet: comparison with others}

It is interesting to compare the non-LTE corrections for 
the O~{\sc i} 7771--5 reported in several published studies with those
estimated following the procedure described in Sect. 3.3 while using 
the same observational equivalent widths along with the same 
$T_{\rm eff}$, $\log g$ and $\xi$ as given in the literature. 
If $W_{\lambda}$ values for two or three lines of the triplet are 
available, we computed the corrections for each of the lines and 
their average was used for the comparison.

\subsubsection{Takeda et al. (1998, 2000)}

Comparisons with the results of our own previous work, 
Takeda et al. (1998, 2000), are shown in Figs. 3a and b. 

The non-LTE corrections for O~{\sc i} 7771.94 given in Table 2 of
Takeda et al.'s (1998) population I G--K giants analysis are in 
good agreement with those derived from our approximate formula 
[Eq. (1)], in spite of the fact that extrapolating Table 3 
was necessary for stars with (4250 K $\la )T_{\rm eff} < 4500$ K.

We can also see that the $\Delta$ values of O~{\sc i} 7771--5 for 
very metal-poor giants computed by Takeda et al. (2000) are 
reasonably consistent with those estimated by the procedure proposed 
in this study. Note, however, that a somewhat large discrepancy is
observed in the $\xi = 5.1$ \kms\ case for HD 184266, where two
different $\xi$ values were suggested depending on the lines
(O~{\sc i} triplet lines and Fe I lines) used for its determination.
This indicates that we should be careful at extrapolating Table 3
to considerably outside of the range they cover (e.g., 
1 \kms $\leq \xi \leq$ 3 \kms ). 

Generally speaking, the results (for O~{\sc i} lines) of our new 
calculations reported 
in this paper (cf. Table 2) are essentially equivalent to our 
previous studies, in spite of the use of updated new atomic model 
in connection with the [O~{\sc i}] transition (cf. Sect. 2.1). 
This can be verified also by comparing
Table 2 in this paper with Table 7 of Takeda (1994), both of which are 
practically equivalent. 
Taking this fact into consideration, we can again confirm from the
coincidence shown in Figs. 3a and b that Eq. (1) 
with Table 3 is sufficiently accurate and useful for 
practical applications.

\subsubsection{Tomkin et al. (1992)}

In Tomkin et al.'s (1992) extensive analysis of C and O abundances
in (from mildly to very) metal-deficient dwarfs, they performed
non-LTE calculations by using a 15 levels/22 lines atomic model
while adjusting the H~{\sc i} collision rate by the requirement that 
the solar O~{\sc i} 7771--5 lines yield the abundance of 8.92 
with the solar model atmosphere of Holweger \& M{\"u}ller (1974). 
As shown in Fig. 4a, the extents of their $|\Delta|$ 
values ($\la 0.1$ dex) tend to be systematically smaller than ours.

\subsubsection{Mishenina et al. (2000)}

Comparing our $\Delta$ values with the recent non-LTE calculation
of Mishenina et al. (2000), who used an atomic model comprising
75 levels/46 transitions and Steenbock \& Holweger's (1984) H~{\sc i} collision
formula with a reduction factor of 1/3, we can see that both are in 
reasonably good agreement in spite of a mild difference in the treatment 
of H~{\sc i} collision rates (cf. Fig. 4b).

\subsubsection{Nissen et al. (2002)}

Nissen et al. (2002) recently carried out a careful abundance study
based on their high-quality data of O~{\sc i} and [O~{\sc i}] lines
(along with OH lines), and obtained almost consistent results
between [O~{\sc i}] (LTE) and O~{\sc i} (non-LTE) at a 
``quasi-linearly increasing'' trend of [O/Fe] within the framework
of the 1D atmosphere analysis (though an application of the 3D correction
to the [O~{\sc i}] abundance causes some discrepancy).
Their O~{\sc i} non-LTE calculation is based on an atomic model
comprising 22 levels/43 transitions, while completely neglecting
the H~{\sc i} collisions. We can see from Fig. 4c that the extent of
their $|\Delta|$ values tends to be larger than ours by $\la 0.15$ dex,
which may be understood as being due to the different treatment of
H~{\sc i} collisions, since neglecting this effect increases $|\Delta|$
by this amount (cf. Appendix A). Even with this moderate difference, 
the reanalysis of their $W_{\lambda}$(O~{\sc i}) and
$W_{\lambda}$([O~{\sc i}]) data based on our $\Delta$ values nearly
reproduces the results they obtained (i.e., the similar tendency for 
O~{\sc i} and [O~{\sc i}]), as shown in Fig. 8a.

\subsubsection{Carretta et al. (2000)}

Carretta et al. (2000) also obtained a nearly consistent solution
of [O/Fe] between O~{\sc i} and [O~{\sc i}] in the metal-poor regime,
but at a ``nearly flat'' trend in contrast to the results of Nissen et al. 
(2002) mentioned above. Namely, they reanalyzed the published 
$W_{\lambda}$(O~{\sc i} 7771--5) data of Tomkin et al. (1992) and 
Edvardsson et al. (1993) and showed that their 
(NLTE-corrected) O~{\sc i} abundances are consistent with those of [O~{\sc i}] 
resulting from their reanalysis of $W_{\lambda}$([O~{\sc i}] 6300/6363) 
data of Sneden et al. (1991) and Kraft et al. (1992).
Their corrections are based on Gratton et al.'s (1999) 
NLTE calculations, where they adopted a model oxygen atom of 
13 levels/28 transitions and a mild enhancement factor of $\sim 3$ 
(determined by the requirement of O~{\sc i} 7771--5/O~{\sc i} 6158 consistency
in RR Lyr variables) applied to classical H~{\sc i} collision rates.

In Fig. 5a, their NLTE corrections 
are compared with our corresponding $\Delta$ values, which were 
evaluated from the $W_{\lambda}$(O~{\sc i} 7771--5) data of Tomkin et al. 
(1992) and Edvardsson et al. (1993) while adopting the $T_{\rm eff}$ and 
$\log g$ values used by Carretta et al. (2000) to make the comparison
as consistent as possible. An interesting tendency is observed in this
figure. That is, while the extents of their non-LTE corrections for 
Edvardsson et al.'s (1993) F--G disk dwarfs are systematically smaller 
than ours, this trend disappears and a good agreement is seen in the 
case of Tomkin et al.'s (1992) considerably metal-poor halo dwarf stars 
(their $|\Delta|$ values are even larger for two stars).

A closer inspection revealed that this is nothing but the dependence
on the metallicity, as shown in Fig. 5b. Namely, their corrections (in terms 
of the absolute $|\Delta|$ values) are smaller than ours for mildly 
metal-deficient
disk stars ($-1 \la$ [Fe/H] $\la 0$), nearly the same for typical halo stars
($-2 \la$ [Fe/H] $\la -1$), and eventually become larger than our values 
for very metal-poor halo stars ($-3 \la$ [Fe/H] $\la -2$).
Evidently, this must be the reason for their ``quasi-flat'' trend of [O/Fe]
obtained for the O~{\sc i} lines, since the extent of their (negative)
non-LTE correction becomes progressively larger than ours (in the relative
sense) as the metallicity decreases, thus suppressing the gradient of 
the [O/Fe] vs. [Fe/H] relation. 

\subsection{Does the metallicity affect the formation of O~I lines?}

In Sect. 4.1.5 we found a metallicity-dependent systematic discrepancy 
between Carretta et al.'s (2000) non-LTE corrections and ours
(i.e., theirs are systematically overcorrected toward a lowered 
metallicity compared to ours),
which must be a very important problem because of its large impact on the
[O/Fe] vs. [Fe/H] relation derived from the O~{\sc i} triplet lines.
Hence, a detailed discussion on this matter
from the viewpoint of the line-formation mechanism may be due.

Actually, such a [Fe/H]-dependent tendency of their correction is clearly 
observed by examining Table 10 of Gratton et al. (1999). 
We plotted their $\Delta$ vs. $W_{\lambda}$(7771.94) relations 
computed for the $\log g = 3.0$ models of different metallicity in 
Figs. 6a and b, each corresponding to $T_{\rm eff}$ = 5000 K 
and 6000 K, where our computed results (Table 2) along with 
the corresponding analytical approximation curves [i.e., Eq. (1) 
with the coefficients taken from Table 3] are overplotted.
It can be seen from these figures that the strong 
metallicity-dependence mentioned above is evidently observed in their
results (especially for the case of $T_{\rm eff}$ = 6000 K).
However, such a tendency (i.e., $\Delta$ is considerably dependent 
on the metallicity at the same $W_{\lambda}$) is not seen 
in our case, where $\Delta$ is nearly a monotonic function
of $W_{\lambda}$ as a first approximation (see in Figs. 6a and b that 
the small symbols of different type continuously follow a common curve 
with a change in $W_{\lambda}$).

In order to demonstrate this more clearly, additional test non-LTE 
calculations were performed for the $T_{\rm eff}$ = 6000 K and
$\log g = 3.0$ models of different metallicities
($[X] = 0, -1, -2, -3$) where the assigned oxygen abundances were
carefully adjusted so as to yield the same (non-LTE) O~{\sc i} 7771.94
equivalent widths for the two metallicity pairs,
$W_{\lambda}$ = 152 m$\rm\AA$ (for $[X]$ = 0 and $-1$) and
$W_{\lambda}$ =  20 m$\rm\AA$ (for $[X]$ = $-2$ and $-3$)
[which were chosen to keep consistency with Gratton et al.'s (1999)
calculation],
and computed the non-LTE correction for each case.
The results are given in Table 4, where the corresponding corrections
taken from Gratton et al.'s (1999) Table 10 are also quoted for a 
comparison. It is clear from this table that such a marked 
metallicity-dependence observed in Gratton et al.'s (1999) corrections
(column 12) is absent in our $\Delta$ values (column 11).

The reason for this characteristics in our case (i.e., near-equality of 
$\Delta$ at the same $W_{\lambda}$ but different metallicities) is 
manifestly explained by Fig. 7, where the behaviors of the source function,
the NLTE-to-LTE line-opacity ratio, and the Planck function
for the four test cases of Table 4 are depicted. We should
note the following facts which can be read from this figure:\\
--- (1) The line opacity for this O~{\sc i} line is nearly in LTE
($l_{0}^{\rm NLTE}/l_{0}^{\rm LTE} \simeq 1$) 
in the line-forming region.\footnote{This is closely related to the fact
that oxygen atoms are dominantly in the neutral stage in the line-forming
region within the framework of our classical model atmosphere, where 
LTE essentially holds in the O~{\sc i}/O~{\sc ii} ionization equilibrium
due to the efficient charge-exchange reaction with hydrogen atoms
(whose population/ionization is computed on the assumption of LTE).
[For example, even in the high-temperature, low-gravity, very metal-poor
atmosphere of ($T_{\rm eff}$ = 6500~K, $\log g = 1$, [Fe/H] = $-3$)
where the ionization is considered to be most important among our
studied cases, $\sim$~99\% of oxygen is neutral at $\tau_{5000} \sim 0.1$.]
Accordingly, the lower-level population of the O~{\sc i} triplet
is significantly affected by the population of the ground level
where most of the O~{\sc i} atoms populate and LTE almost prevails.
Note that the photoionizing radiation field (which is conspicuously
dependent on the metallicity via the line/continuum opacities)
does not play any important role for the O~{\sc i} population.}
Though an appreciable departure from LTE (i.e., overpopulation) is
observed at the shallower layer especially for the metal-poor 
atmospheres, it does not have any significant affection upon 
the emergent radiation. [See Sect. 4.2 in Takeda (1994) for the
explanations on these characteristics of $l_{0}^{\rm NLTE}/l_{0}^{\rm LTE}$.]\\
--- (2) As can be seen from Fig. 7, the line source function ($S_{\rm L}$)
of this O~{\sc i} 7771.94 line resulting from our detailed non-LTE
calculations (thin solid lines) is essentially the same as the 
two-level-atom-type line source function (open circles). Namely, 
the solution obtained by solving the classical
scattering problem with the simple form of $S_{\rm L}^{\rm two-level}$,
\begin{equation}
S_{\rm L}^{\rm two-level} = (\overline{J} + \epsilon' B)/(1 + \epsilon'),
\end{equation}
while using the realistic values of $\epsilon'(\tau)$, 
$B(\tau)$, and $\eta(\tau)$ computed from the model atmospheres 
[where $\overline{J}$ (frequency-averaged mean radiation field), 
$\epsilon'$ (collisional-to-radiative ratio of the deexcitation rate),
$B$ (Planck function), and $\eta(\tau)$ (the line-to-continuum opacity 
ratio necessary to compute the line optical depth and to combine the 
line-and-continuum source function into the total source function)
are the quantities related to the transition between the two levels; 
cf. Eq. (11-6) in Mihalas (1978) or Appendix B in this paper], yields 
essentially the same result as $S_{\rm L}^{\rm true}$, obtained from 
the solutions of detailed statistical-equilibrium calculations, 
\begin{equation}
S_{\rm L}^{\rm true} = 
(2 h \nu^{3}/c^{2}) [(n_{i}g_{j})/(n_{j}g_{i}) - 1]^{-1},
\end{equation}
where $n_{i}$/$n_{j}$ and $g_{i}$/$g_{j}$ are the number population
and the statistical weight of lower/upper level, respectively
[cf. Eq. (4-14) in Mihalas (1978)].
This evidently suggests that {\it the formation 
of the O~I triplet is essentially determined by the quantities
directly related to this transition itself} without any significant 
relevance to the details of the adopted atomic model.\footnote{
The line source function of any transition [given by Eq. (3)]
can be generally expressed (by rewriting the rate equations) as
$S_{\rm L}^{\rm true} = 
(\overline{J} + \epsilon' B + c_{1})/(1 + \epsilon' + c_{2})$,
where $c_{1}$ and $c_{2}$ are the interlocking terms representing
the cumulative influences of other levels and transitions
[cf. Eq. (12-4) in Mihalas (1978)]. Then, the consequence of 
$S_{\rm L}^{\rm true} \simeq S_{\rm L}^{\rm two-level}$
naturally implies that the complicated effects of other levels and 
transitions (represented by $c_{1}$ and $c_{2}$) are almost negligible; 
i.e., these O~{\sc i} 7771--5 lines form as if the oxygen atom were 
made of only the two levels relevant to this transition.}\\
--- (3) In this case, as the well-known characteristics of such a classical 
``two-level-atom'' problem (e.g., Hummer 1968; Appendix B), the extent 
in the dilution of the line source function ($S_{\rm L}/B$) 
is controlled by the line-to-continuum opacity ratio 
$\eta$ ($\equiv l_{0}/\kappa$). 
And since both eventually determine the strength of the line, it is quite
natural that the non-LTE effect (i.e., the extent of the $S_{\rm L}$ 
dilution) is closely coupled with the line strength ($W_{\lambda}$).

Then, there is almost no room for the metallicity directly playing
any significant role in the non-LTE effect. 
Admittedly, there are some possibilities of metallicity-dependence 
even within the framework of such a two-level-type line-formation. 
However, any of them are unlikely in the present case. 
For example, appreciable [Fe/H]-dependences of $T$ (metallicity-related 
surface cooling) or of $l_{0}^{\rm NLTE}/l_{0}^{\rm LTE}$ 
[see (1) above] existing near to the surface (cf. Fig. 7) do not matter, 
because they occur far above the line-forming region.
Also, the collisional rate between two levels (affecting the photon 
destruction probability $\epsilon'$) is hardly dependent on the metallicity 
(or electron density) in our case because of the dominance 
of the H~{\sc i} collisions\footnote{For example, the values of 
$C_{\rm H}/C_{\rm e}$ (at $\tau_{5000} \sim 0.1$) for the O~{\sc i} 7771--5
transition, computed from Steenbock \&Holweger's (1984) formula (for 
$C_{\rm H}$) and Van Regemorter's (1962) formula (for $C_{\rm e}$)
(as adopted in this study) for each of the representative models
with parameters of ($T_{\rm eff}$, $\log g$, [Fe/H]) are as follows:
$\sim$30/$\sim$1000 for (4500~K, 4.0, 0/$-2$),
$\sim$8/$\sim$500 for (4500~K, 1.0, 0/$-2$),
$\sim$7/$\sim$100 for (5500~K, 4.0, 0/$-2$),
$\sim$4/$\sim$10 for (5500~K, 1.0, 0/$-2$),
$\sim$2/$\sim$6 for (6500~K, 4.0, 0/$-2$), and
$\sim$0.3/$\sim$0.5 for (6500~K, 1.0, 0/$-2$).
Hence, except for the high-temperature and low-gravity model (which is 
almost irrelevant in the present study of metal-poor stars), 
$C_{\rm H}$ is generally predominant over $C_{\rm e}$.}
 [though an extreme case of pure electron
collisions (neglecting H~{\sc i} collisions) would lead to 
a metallicity-dependence of the $\epsilon'$ values (and eventually 
of the non-LTE effect) owing to the role of metallicity as 
the electron donor; e.g., Kiselman (1991)].

A supplementary discussion based on the two-level-atom along with
the simple Milne--Eddington model is given in Appendix B, where
the $T_{\rm eff}$/$\log g$-dependence of the $\Delta$ vs. $W_{\lambda}$
relation given in Fig. 1 (along with its [Fe/H]-independence) is 
qualitatively explained more in detail.

Accordingly, the results of the Padova group, that the extent
of the non-LTE correction increases with a decrease in the 
metallicity (for a given $W_{\lambda}$), is hard to understand 
at least from the viewpoint of what we found from our calculations.
Namely, if their results are to be seriously taken,
something in their calculations must be greatly different from ours
(e.g., breakdown of the two-level-atom nature and the dominant
contributions from other levels and transitions).
Curiously, however, an inspection of Gratton et al.'s (1999) Fig. 6 suggests
that their non-LTE departure coefficients behave themselves consistently
with our computational results (see, e.g., Fig. 9 of Takeda 1994); that is, 
for the case of $T_{\rm eff}$ = 6000 K and $\log g$ = 4.5, the departure 
from LTE is larger for the [Fe/H] = 0 model than for the [Fe/H] = $-2$ model.
In other words, their non-LTE calculations appears quite reasonable
in this respect. Then, how could their calculations lead to the tendency 
of enhanced $|\Delta|$ with a lowered [Fe/H] ? 
Anyway, it appears that the non-LTE corrections given in
Gratton et al.'s (1999) Table 10 and those used by Carretta et al. (2000) 
are questionable, and thus their consequences had better be viewed 
with caution. This criticism also applies to any other results 
obtained by using their corrections; e.g., the ``flat-like'' [O/Fe] tendency 
derived by Primas et al. (2001).

\subsection{Validity of LTE for the [O~I] 6300/6363 lines}

As mentioned in Sect. 1, the search for the possibility of the non-LTE
effect in the [O~{\sc i}] lines, based on the new (more realistic) 
O~{\sc i} atom, was actually one of the main aims in this paper. 

According to the present calculation, however, the departure coefficients
of the lower term (2s$^{2}$~2p$^{4}$ $^{3}$P; hereinafter referred to 
as ``term 1'') and the upper term (2p$^{4}$~$^{1}$D; hereinafter ``term 2'')
for the [O~{\sc i}] 6300/6363 lines turned out to be essentially equal to 
unity over almost all atmospheric layers, which guarantees the validity of
the assumption of LTE for these forbidden [O~{\sc i}] lines. 

The reason for this is attributed to the fact that (1) the population of
the lower ground term is thermalized relative to the continuum (i.e.,
accomplishment of LTE population) owing to the efficient charge-exchange
reaction with H atoms, and (2) these lower and upper terms are strongly 
coupled with collisions which equalize the $b$ values of these two terms. 

It was further found after several test calculations that this is a rather 
robust conclusion, which may not be affected by any uncertainty in 
the adopted collisional cross section.
That is, although little is known and considerable ambiguities are involved
in the collisional rates due to H~{\sc i} atoms for this 1--2 transition 
($C_{1-2}^{\rm H}$; for which we used the scaled value estimated from 
$C_{1-2}^{\rm e}$), we obtained essentially the same results {\it even if 
$C_{1-2}^{\rm H}$ was completely neglected}; i.e., only the $C_{1-2}^{\rm e}$ 
itself is sufficiently large to bring the $b$ vales of terms 1 and 2 
into the same value of unity. 

We also carried out a simulation for the limiting test case of completely
neglecting the collisional interaction between terms 1 and 2
($C_{1-2}^{\rm e} = C_{1-2}^{\rm H} = 0$). In this case, an appreciable
overpopulation in the upper term 2 was observed (the extent of its NLTE 
departure growing toward the shallower layer) as a result of the downward 
cascade via the UV transitions connecting to term 2, which makes the NLTE 
line strength somewhat smaller compared to the LTE value (i.e., a positive
correction in terms of $\Delta$ defined in Sect. 3.1) because of the
resulting inequality of $S_{\rm L}/B > 1$ in the upper layer.
Even in such an extreme case, however, the extent of the computed
$\Delta$ value is $\la 0.05$~dex in most range of the atmospheric parameters
and definitely insignificant in the practical sense.
Accordingly, we can confidently conclude that the non-LTE effect can be
safely neglected for the [O~{\sc i}] lines, for which LTE must be a valid
approximation.

\subsection{[O/Fe] vs. [Fe/H] relation}

\subsubsection{Procedure of reanalysis}

Now we are ready to study the [O/Fe] vs. [Fe/H] relation
of metal-poor stars by reanalyzing the published equivalent-width
data of O~{\sc i} 7771/7774/7775 and [O~{\sc i}] 6300/6363 lines 
while applying the non-LTE corrections based on our calculations.

We adopted the same $T_{\rm eff}$, $\log g$, [Fe/H], and $\xi$ values
as those used in the literature, from which the data of equivalent widths
were taken. 
Regarding the model atmospheres, we used Kurucz's (1993)
grid of ATLAS9 models ($\xi$ = 2 \kms ) which were interpolated 
with respect to $T_{\rm eff}$, $\log g$, and [Fe/H] of a star.
As in Sect. 3.2, Kurucz's WIDTH9 program was invoked for determining
the LTE abundance ($\log\epsilon_{\rm O}^{\rm LTE}$) 
while using the line data given in Table 1.
Then, the [O/Fe] ratio is obtained as
\begin{equation}
[{\rm O}/{\rm Fe}] \equiv 
( \log\epsilon_{\rm O}^{\rm LTE} + \Delta - 8.93) - [{\rm Fe}/{\rm H}] ,
\end{equation}
where $\Delta$ is the non-LTE abundance correction, which was evaluated 
from $W_{\lambda}$, $T_{\rm eff}$, $\log g$, and $\xi$ 
based on the procedure described in Sect. 3.3 [i.e., application of
Eq. (1) with Table 3] for the case of O~{\sc i}~7771/7774/7775 lines,
and $\Delta = 0$ was adopted for the [O~{\sc i}]~ 6300/6363 lines
(cf. Sect. 4.3).
Note that the solar oxygen abundance was assumed to be 8.93 (in the 
usual scale of $\log\epsilon_{\rm H} = 12$) according to Anders \& Grevesse 
(1989).
We treat [O/Fe] derived from O~{\sc i} 7771--5 lines and [O~{\sc i}] 6300/6363 lines
separately, which we hereinafter referred to as
[O/Fe]$_{77}$ and [O/Fe]$_{63}$, respectively.
In case that equivalent-width data are available for two or three O~{\sc i} 
triplet lines, or for both [O~{\sc i}] lines, we calculated [O/Fe] 
for each line and adopted their simple mean to obtain [O/Fe]$_{77}$ or 
[O/Fe]$_{63}$.

\subsubsection{Results of [O/Fe]}

We tried to use as many $W_{\lambda}$ data as possible taken from various 
sources, though our literature survey is not complete.
Consequently,  $\sim 300$ stars ($\sim 200$ for [O/Fe]$_{77}$
and $\sim 100$ for [O/Fe]$_{63}$) covering a metallicity range of
$-3 \la$ [Fe/H] $\la 0$ could be analyzed, which may be sufficient for
statistically meaningful discussion.
The adopted data of the atmospheric parameters ($T_{\rm eff}$, $\log g$,
$\xi$) the equivalent widths for each star are given in Table 5 
(available only in electronic form), where the results of the 
LTE abundances and the non-LTE corrections/abundances are also
presented.
The resulting [O/Fe] ratios (i.e., [O/Fe]$_{77}$ or [O/Fe]$_{63}$) 
plotted with respect to [Fe/H] for each star are shown in Fig. 8a,
and the corresponding non-LTE corrections applied to the O~{\sc i} 7771--5
lines are also plotted in Fig. 8b.
The mean $\langle[$O/Fe$]_{77}\rangle$ and $\langle[$O/Fe$]_{63}\rangle$, averaged in 
each [Fe/H] bin of 0.5 dex, are given in Table 6.
The spline-fit curves for the runs of $\langle[$O/Fe$]_{77}\rangle$ and 
$\langle[$O/Fe$]_{63}\rangle$ with [Fe/H] are also depicted in Fig. 8a.
Our conclusions read from Fig. 8 and Table 6 are as follows.

(1) For mildly metal-deficient disk stars ($-1 \la$ [Fe/H] $\la 0$), 
our [O/Fe]$_{77}$ and [O/Fe]$_{63}$ show a quite similar 
behavior; i.e., a gradual increase from [O/Fe]$\sim 0$ ([Fe/H] $\sim 0$) 
to [O/Fe]$\sim +0.5$ ([Fe/H] $\sim -1$). This means that our non-LTE 
corrections are successful to achieve an agreement between the
oxygen abundances derived from [O~{\sc i}] 6300/6363 and O~{\sc i} 7771--5 for 
those population I stars, which is consistent with the conclusion
of Takeda et al. (1998).

(2) However, a discrepancy begins to appear when we go toward a lower
metallicity further down from [Fe/H] $\sim -1$. That is, [O/Fe]$_{77}$
still increases with a slight change (i.e., gentler) of the slope 
from [O/Fe]$\sim +0.5$ ([Fe/H] $\sim -1$) 
to [O/Fe]$\sim +1$ ([Fe/H] $\sim -3$), while [O/Fe]$_{63}$ maintains
nearly the same value at [O/Fe]$\sim +0.5$ on the average 
over the metallicity range of $-2.5 \la$ [Fe/H] $\la -1$ showing
a quasi-flat plateau. This kind of [O/Fe]$_{77}$ vs. [O/Fe]$_{63}$ discordance 
for those conspicuously metal-poor halo stars is just what has been 
occasionally reported so far (cf. Sect. 1).
Our analysis thus suggests that this discrepancy in halo stars (amounting to 
$\sim$ 0.2--0.3 dex at $-2.5 \la$ [Fe/H] $\la -1.5$) {\it does} exist,
without being successfully removed by our non-LTE corrections,
unlike the conclusion of Carretta et al. (2000).
We will discuss this problem more in detail in the next section.

\subsection{O~I vs. [O~I] discrepancy}

\subsubsection{Discordance in halo stars}

We have thus confirmed the existence of discrepancy between
[O/Fe]$_{77}$ and [O/Fe]$_{63}$ in the quite metal-poor regime 
($-3 \la$ [Fe/H] $\la -1$), while a good agreement was obtained
in disk stars ($-1 \la$ [Fe/H] $\la 0$).\footnote{Rather 
exceptionally, the recent data of Nissen et al. (2002) suggest 
more or less consistent abundances for both O~{\sc i} and 
[O~{\sc i}] (cf. double squares and circled pluses in Fig. 8)
even for the very metal-poor stars as they confirmed in their 
1D analysis. We will return to this point in Sect. 4.5.2.}

This can not be successfully removed by the non-LTE effect
though they surely act in the direction of mitigating the discordance.
According to our opinion (cf. Sect. 4.2), the O~{\sc i} vs. [O~{\sc i}] 
consistency obtained by Carretta et al. (2000) may be nothing but a 
fortuitous coincidence caused by their questionable non-LTE corrections
for the O~{\sc i} 7771--5 triplet (i.e., systematically overcorrected 
toward a lower metallicity).

It is also unlikely that systematic errors in the $T_{\rm eff}$ scale
(affecting mainly on O~{\sc i}) is responsible for the cause of 
this discrepancy as has been occasionally suspected (e.g., King 1993), 
since our analysis in Sect. 4.4 is based on the mixed data of 
equivalent widths and atmospheric parameters taken from a variety of 
data sources.

Hence, we would consider that this discordance in halo stars is more or 
less inevitable as far as the conventional abundance determination method 
is concerned (i.e.,  ``standard'' non-LTE calculations as well as
model atmospheres); that is, a challengingly novel modelling would be
required to bring them into agreement.\footnote{More realistic 
line-formation calculation based on 3D inhomogeneous model atmospheres
might be a candidate. However, to our current knowledge, this effect
does not improve the situation; rather, it even increases the discrepancy
(i.e., the O~{\sc i} abundance is barely affected, while the 
[O~{\sc i}] abundance is somewhat {\it lowered} (cf. footnote 1).}

\subsubsection{Search for a trend in terms of $T_{\rm eff}$ and $\log g$}

Then, the next question of interest is naturally 
``which solution is correct ?'', or in other words 
``which yields the wrong abundances, O~{\sc i} 7771--5 or 
[O~{\sc i}] 6300/6363 ?''. While this is a difficult question,
it may be worthwhile as a first step to study the nature of the
discordance; i.e., its dependence on $T_{\rm eff}$ and $\log g$.

Let us invoke the following guiding principle:
If we assume that the origin of the error is concerned with 
some stellar physical property which is not appropriately modelled, 
it is reasonable to expect that abundance discrepancies shown by 
the erroneous side (whichever O~{\sc i} or [O~{\sc i}]) may show 
some kind of systematic tendency in terms of these stellar parameters, 
since any stellar physical environment must undergo 
a distinct change if these parameters vary substantially.
And, in an ideal case, the discordance might asymptotically disappear 
if we go to either limit of the parameter variation.

Motivated by this thought, we paid attention to 
[O/Fe]$_{77} - \langle[$O/Fe$]_{63}\rangle$ (the discrepancy of 
{\it each star's} [O/Fe]$_{77}$ relative to the {\it mean} $\langle[$O/Fe$]_{63}\rangle$)
and [O/Fe]$_{63} - \langle[$O/Fe$]_{77}\rangle$ (the inverse of the above)
within two metallicity bins of $-2.75 \leq$ [Fe/H] $< -2.25$
and $-2.25 \leq$ [Fe/H] $< -1.75$, where a clear discordance of
$\langle[$O/Fe]$_{77}\rangle - \langle[$O/Fe$]_{63}\rangle \sim 0.3$ dex is observed (cf. Table 6).
The dependence of former quantity for each stars is plotted in Fig. 9
with respect to $T_{\rm eff}$ (a) and $\log g$ (b), while that of the latter
is similarly shown in Fig. 10.

In Fig. 9, what we observe is mostly a large scatter around 
the mean of $\sim +0.3$, which is presumably caused by the strong 
$T_{\rm eff}$-sensitivity of these high-excitation O~{\sc i} 7771--5 lines. 
However, we can recognize in Fig 9a a marginal trend that
[O/Fe]$_{77}- \langle[$O/Fe$]_{63}\rangle$ tends to
decrease toward higher $T_{\rm eff}$ (especially at $T_{\rm eff} \ga 6000$ K).
Also, if we pay attention to the data for $-2.75 \leq$ [Fe/H] $< -2.25$ 
in Fig. 9b (open circles), a slight $\log g$-dependence appears to exist 
in the sense that higher-gravity stars tend to show less extent of 
the discordance.
Similarly, Fig. 10 also suggests a weak trend for this group of 
very metal-poor stars (open circles); 
i.e., $|$[O/Fe]$_{63} - \langle[$O/Fe$]_{77}\rangle|$ appears to decrease 
with an increase in $T_{\rm eff}$ as well as in $\log g$ toward 
the high-$T_{\rm eff}$/high-$\log g$ limit.
Consequently, a weak systematic trend may exist, especially for
the very metal-deficient case of $-2.75 \leq$ [Fe/H] $< -2.25$,
in the sense that the discrepancy tends to be lessened for an increase
in $T_{\rm eff}$ as well as in $\log g$.

However, these suspected $T_{\rm eff}$- and $\log g$-dependences
should not be regarded as being independent. We should recall
that the very metal-poor stars ([Fe/H] $\la -2$) analyzed so far
are mainly giants or subgiants. Namely, since $T_{\rm eff}$ and 
$\log g$ show a close positive correlation in metal-poor giants
reflecting the position on the HR diagram (see, e.g., Fig. 3 of
Pilachowski et al. 1996), the rather similar behavior of
the O~{\sc i} $-$ [O~{\sc i}] discrepancy in terms of the
changes in both $T_{\rm eff}$ and $\log g$ may be naturally
understood.

Consequently, what we can learn from Figs. 9 and 10 is,
modestly speaking, that analyzing higher $T_{\rm eff}/\log g$
subgiants (and maybe also dwarfs) has a less probability of
suffering a large O~{\sc i} $-$ [O~{\sc i}] discrepancy
as compared to the study of lower $T_{\rm eff}/\log g$ giants.
The recent results of Nissen et al. (2001) (cf. footnote 10)
may be regarded as supporting this speculation, since their data
are based on stars with $\log g > 3$.

Unfortunately, however, Figs. 9 and 10 do not allow us any 
definitive arguments regarding the cause of the discrepancy or 
which of the O~{\sc i} or [O~{\sc i}] is problematic.
It is apparent from these figures that the [O~{\sc i}] data of 
very metal-poor subgiants and dwarfs with higher $T_{\rm eff}/\log g$ 
(and inversely the O~{\sc i} data for very metal-poor giants with lower 
$T_{\rm eff}/\log g$) are quite insufficient, reflecting the preference
of [O~{\sc i}] lines for giants so far (because their strengths increase 
with a lowering of the gravity).
Hence, in order for a reliable statistical discussion, 
much more observations of [O~{\sc i}] lines for very metal-poor
dwarfs or subgiants of higher gravity (along with O~{\sc i} lines 
for very metal-poor giants of lower gravity) are desirably awaited,
which would surely improve this situation toward clarifying the origin
of the discrepancy.

\section{Conclusion}

For the purpose of investigating the line formation of
O~{\sc i} 7771--5 lines of IR triplet and [O~{\sc i}] 6300/6363 lines
especially concerning the consistency of the oxygen abundances
derived from these two in metal-poor disk/halo stars,
we carried out non-LTE calculations with a new atomic model of oxygen
on an extensive grid of model atmospheres 
(4500~K $\leq T_{\rm eff} \leq $ 6500~K, $1 \leq \log g \leq 5$,
and $-3 \leq [{\rm Fe}/{\rm H}] \leq 0$).

In order to make the resulting non-LTE abundance corrections of 
O~{\sc i} 7771--5 lines more useful for practical applications,
we derived an analytical formula with appropriate coefficients
computed and tabulated in a grid of atmospheric parameters,
by which the non-LTE correction is easily evaluated for any given set of
$W_{\lambda}$, $T_{\rm eff}$, $\log g$ and $\xi$.

We discussed our non-LTE corrections for the O~{\sc i} triplet lines while 
comparing them with those derived from other studies published so far.
We could not reproduce the trend of the non-LTE correction derived 
by Carretta et al. (2000), the systematic [Fe/H]-dependence of which 
is essentially the reason for their success of accomplishing the 
O~{\sc i} vs. [O~{\sc i}] consistency.

In contrast, we have shown that the non-LTE effect of O~{\sc i} 7771--5
lines is well described by the classical two-level-atom model
and the metallicity can not be an essential factor. 
As a matter of fact, the formation of this O~{\sc i} triplet was found
to be quite simple in the sense that only the transition-related quantities
are important and details of the atomic model are not very significant,
supporting the arguments of Kiselman (2001) and Nissen et al. (2002)
who also arrived at the same conclusion.

Regarding the forbidden [O~{\sc i}] lines, we arrived at a robust 
conclusion that the non-LTE effect for the [O~{\sc i}] 6300/6363 lines 
can be safely negligible and thus LTE is a valid approximation.

We reanalyzed the various published equivalent-width data of 
O~{\sc i} 7771--5 and [O~{\sc i}] 6300/6363 lines (while the non-LTE
effect is taken into account for the former based on our calculations), 
in order to study the behavior of [O/Fe] vs. [Fe/H] relation
over a wide range of metallicity ($-3 \la [{\rm Fe}/{\rm H}] \la 0$).
While [O/Fe]$_{77}$ and [O/Fe]$_{63}$ were found to agree with each other 
for disk stars ($-1 \la [{\rm Fe}/{\rm H}] \la 0$), the existence of
a systematic discrepancy was confirmed for halo stars
($-3 \la [{\rm Fe}/{\rm H}] \la -1$) amounting to 
[O/Fe]$_{77} - $ [O/Fe]$_{63} \sim 0.3$ dex at 
$-3 \la [{\rm Fe}/{\rm H}] \la -2$.

An inspection of the dependence of discrepancy upon $T_{\rm eff}$ and 
$\log g$  suggested that the extent of the discrepancy tends to be
comparatively lessened for higher $T_{\rm eff}/\log g$ stars,
indicating the preference of dwarf (or subgiant) stars for 
studying the oxygen abundance of metal-deficient stars.

\appendix

\section{On the effect and treatment of H~{\sc i} collision}

The inelastic collision rate due to neutral hydrogen atoms
is regarded to be one of the important ambiguous factors in 
non-LTE calculations on late-type stellar atmospheres.
A rough classical formula derived by Steenbock and Holweger (1984)
is available for this (see Takeda 1991 for more detailed description 
on the standard classical recipe adopted).
However, because of its nature of only an order-of-magnitude accuracy,
a correction factor is often introduced, by which this classical rate 
is to be multiplied. 
We express this factor as $10^{h}$ by using the logarithmic correction ($h$). 
Negative $h$ leads to the suppression of the H~{\sc i} collisional rates
compared to the classical case, which increases the extent of 
the NLTE abundance correction ($|\Delta|$); and vice versa.
This situation is depicted in Fig. 11, where the $h$-dependence
of $\Delta$ for the O~{\sc i} 7774.18 line is shown,
which was calculated for selected representative models.

As mentioned in Sect. 2.1, we used $h = 0$ (as was done by Takeda 1992, 1994)
throughout this paper, which means the simple adoption of 
the classical treatment without any correction.
This choice stems from the detailed investigation of Takeda (1995), 
where $h = 0.0 (\pm 0.5)$ was concluded from the analysis of 
the solar flux spectrum based on (1) the comparison of the solar 
oxygen abundance derived from O~{\sc i} 7771--5 lines with the average 
of other 15 weaker lines and (2) the profile-fitting of these near-IR
triplet lines. 

It may be worthwhile, from another point of view, to check 
the consistency of this treatment on nearby solar-type stars, 
which have been recently analyzed by Takeda et al. (2001) 
to study the abundance characteristics of planet-harboring stars.

Fig. 12 shows the comparison of the non-LTE abundance of 
O~{\sc i} 7771--5 triplet (with the NLTE correction
obtained by using the formula described in this study)
with the LTE abundance of O~{\sc i} 6158.2 (a) and [O~{\sc i}] 6300.3 (b),
which were derived from either equivalent-widths or spectrum-synthesis
technique (with the same line data as given in Table 1) as described in 
Takeda et al. (2001).
Note that LTE can be safely applied to these latter two lines
(cf. Table 7 and Sect. 4.3).

We can see from Fig. 12a that $\log\epsilon_{6158}^{\rm O}$
and $\log\epsilon_{7771-5}^{\rm O}$ are in satisfactory agreement
with each other, which confirms that our choice of $h=0$ is surely
reasonable at least within the framework of our non-LTE calculations.

Meanwhile, for the case of 6300 vs. 7771--5 comparison shown 
in Fig. 12b, the scatter is comparatively large, which makes 
this figure less useful for the purpose of checking the adequacy of $h$.
Presumably, this appreciable spread may be due to the large difference 
in the $T_{\rm eff}$-sensitivity of these two lines (in contrast to
the case of 7771--5 and 6158 which have almost the same 
$T_{\rm eff}$-dependence).
A close inspection of Fig. 12b suggests that the 6300 abundance
tends to be somewhat larger especially for the case of
higher oxygen abundance ($\log\epsilon \ga 9.0$), for which
two reasons may be suspected; i.e., the possible inaccuracy 
in the adopted $gf$ value and/or the effect of blending.
That is, according to the recent study of the solar [O~{\sc i}] 6300
line profile by Allende Prieto et al. (2001), the Ni~{\sc i} line at 
6300.34~$\rm\AA$ makes an appreciable contribution to this feature.
Our neglect of this effect may have cause an overestimation.
In addition, their paper suggests that the most recently calculated
$gf$ value is $\log gf$([O~{\sc i}]~6300) = $-9.72$, which is by +0.1 dex
larger than that adopted in this study (cf. Table 1). 
Hence, our solution of $\log\epsilon_{6300}^{\rm O}$ may have been 
somewhat overestimated also from this point of view. 

We finally remark that, even if our $\log\epsilon_{6300}^{\rm O}$ values 
have been overestimated either due to the use of underestimated 
$\log gf$ or due to the blending effect of the Ni line, our conclusion 
of 6300 vs. 7771--5 abundance discrepancy (cf. Sect. 4.4.2) can not be 
changed at all (i.e., because the discordance would become even larger).

\section{Interpretation of the triplet-line formation with the two-level-atom
model}

It was shown in Sect. 4.2 that the formation of O~{\sc i} 7771--5 lines
can be well described by the ``two-level'' atomic model.
Based on this simple line source function [cf. Eq. (2)] with
the  classical Milne--Eddington model (i.e., depth-independent 
line-parameters with the continuum source function linearly increasing
with $\tau$), we try to explain the results of our non-LTE calculations,
that the NLTE correction for the O~{\sc i} 7773 triplet (for given 
$T_{\rm eff}$, $\log g$, and $\xi$) is nearly a monotonic function of 
$W_{\lambda}$ (irrespective of any reasonable change in the metallicity).
Though this is a very simple schematic model, it gives us a deep insight
on the formation of this O~{\sc i} line.

\subsection{What we can learn from the basic characteristics}

It was Hummer (1968) who first showed that the extent of the dilution 
in the two-level-atom line source function defined by Eq. (2) is 
determined by 
(a) the photon-destruction probability defined as the ratio of the
radiative to collisional (downward) transition probability 
($\epsilon' \simeq C_{21}/A_{21}$; cf. equation [11-7] in Mihalas 1978) and 
(b) the relative importance of the thermalizing continuum 
determined by the line-to-continuum opacity ratio 
$\eta \equiv \kappa_{\rm L}/\kappa_{\rm C}$ (cf. Hummer 1968).\footnote{
Hummer (1968) used a quantity called $\beta$ ($\equiv 
\kappa_{\rm C}/\kappa_{\rm L}$; the continuum-to-line opacity
ratio) for the same purpose, which is just equivalent to $\eta^{-1}$.}
Namely:\\
(i) For the same $\epsilon'$ (i.e., atmospheres with the same 
temperature and density), the extent of $\eta$ essentially controls
the dilution of $S_{\rm L}$ is (i.e. the larger $\eta$,
the more enhanced NLTE effect).\\
(ii) On the other hand, the extent of the dilution in $S_{\rm L}$ 
naturally determines the importance of the NLTE effect (i.e., 
the extent of the NLTE correction) as well as the strength of the
observed line (i.e., equivalent width) in this case of scattering
line formation.\\
We now understand from ``fact (ii)'' that a close relationship exists 
between the strength of the O~{\sc i} 7771--5 triplet lines and 
the corresponding non-LTE corrections. Meanwhile, ``fact (i)'' simply tells
that the essential key parameter is $\eta$ only (for the same $\epsilon'$),
which further suggests that the difference in the metallicity does not 
play any essential role in the present problem.\footnote{While the 
possibilities of ``indirect'' metallicity effects via the temperature 
structure or the collisional rates surely exist, they are insignificant
in the present case, as mentioned in Sect. 4.2.}
Namely, as far as the line-to-continuum opacity ratio is the same 
(along with the same $T_{\rm eff}$ and $\log g$), the same extent of 
$S_{\rm L}$, the same equivalent width, and the same non-LTE correction 
are expected irrespective of the metallicity, thus yielding the same 
$\Delta$ vs. $W_{\lambda}$ correlation.

\subsection{Numerical simulation toward a qualitative explanation of Fig. 1}

It may be instructive to compare the results from a simple simulation
based on such a ''two-level'' model with the actually
obtained $\Delta$ vs. $W_{\lambda}$ relation shown in Fig. 1.
Though the value of $\epsilon'$ is strongly depth-dependent in 
real atmospheres, we found $\epsilon' \sim 1$ (dwarfs) 
and $\epsilon' \sim 0.1$ (supergiants) in the line-forming region
from actual model calculations. We adopted the Milne--Eddington model
assuming the depth-independent line-to-continuum opacity ratio
$\eta$ and the linear form of the Planck function,
$B(\tau) \equiv B_{0} + B_{1}\tau = B_{0}(1 + B_{1}/B_{0}\tau)$ 
($\tau$: continuum optical depth).
We fixed $B_{0} = 1$ and changed the gradient $B_{1} (= B_{1}/B_{0})$.
Again the determination of this gradient at 7773~$\rm\AA$ 
is rather difficult since it is not linear in terms of $\tau$
in actual cases, but we tentatively chose by inspection of 
the model atmospheres $B_{1} = 2$ (low $T_{\rm eff}$ case) and 
$B_{1} = 1$ (high $T_{\rm eff}$ case).
Hence, the radiative transfer problems (two-level-atom,
complete frequency redistribution, and the pure Gaussian line profile)
investigated by Hummer (1968) were numerically solved by using 
the Rybicki's (1971) scheme for various values of $\eta$ 
ranging from $10^{-2}$ to $10^{3}$ for each of the three combinations of 
($B_{1}$, $\epsilon'$); (2, 1), (2, 0.1), and (1, 1). 
And then the emergent flux spectra were computed from the resulting 
solution of the source function (NLTE case), along with the LTE case 
of $S_{\rm L} = B$ (i.e., the limit of $\epsilon' \rightarrow \infty$), 
to obtain the equivalent width ($W$).

In Fig. 13 are shown the LTE and NLTE curves of growth [panel (a)],
and the corresponding NLTE correction $\Delta\log\eta$
($\equiv \log\eta_{\rm NLTE} - \log\eta_{\rm LTE}$; i.e., difference of 
the abscissa for a given $W$) as a function of $W_{\rm NLTE}$ [panel (b)].
We can see from Fig. 13b that the computed 
$\Delta\log\eta$ vs. $W_{\rm NLTE}$ curves well resemble 
(at least qualitatively) the $\Delta$ vs. $W_{\lambda}$
relations observed in Fig. 1, in the sense that 
they can be approximated with the functional form of Eq. (1).

It is worthwhile paying attention to the effects of changing $\epsilon'$ 
or $B_{1}$ observed in Fig. 13b.
As expected, the extent of the NLTE correction increases with a decrease in 
$\epsilon'$. Meanwhile, that $|\Delta\log\eta|$ is larger for a smaller 
gradient ($B_{1}$) of the Planck function may be attributed to the decreased 
sensitivity of $W_{\rm LTE}$ (to a change in $\eta$) owing to
the lessened gradient (i.e., $W_{\rm LTE}$ is relatively more affected 
by the Planck function $B$, in contrast to the NLTE case which is mainly 
controlled by the radiation field).
It should be noted that these two qualitative behaviors concluded from 
our simple model reasonably explain the tendencies observed in Fig. 1. 
That is, the larger $|\Delta|$ for a lowered gravity (at a given 
$W_{\lambda}$) is reasonably explained by the $\epsilon'$-dependence 
described above. Meanwhile, regarding the tendency of larger 
$|\Delta|$ for a higher $T_{\rm eff}$  (at a given $W_{\lambda}$), 
the above-mentioned effect of decreased $B_{1}$ may be responsible for it.

\clearpage

\renewcommand{\thetable}{\arabic{table}}
\renewcommand{\thefigure}{\arabic{figure}}

\clearpage
\setcounter{table}{0}
\begin{table}[h]
\small
\caption{Data of neutral oxygen lines adopted for abundance calculations.}
\begin{center}
\begin{tabular}
{crrrrlc}\hline \hline
RMT & $\lambda$ ($\rm\AA$) & $\chi$ (eV) & $\log gf$ & Gammar & Gammas &
Gammaw \\
\hline
1  & 7771.944 & 9.146 &  0.324 & 7.52 & $-5.55$ & $-7.65^{*}$ \\
1  & 7774.166 & 9.146 &  0.174 & 7.52 & $-5.55$ & $-7.65^{*}$ \\
1  & 7775.388 & 9.146 & $-0.046$ & 7.52 & $-5.55$ & $-7.65^{*}$ \\
$\cdots$ & 6300.304 & 0.000 & $-9.819$ & $-2.17$ & $-7.83^{*}$ & $-8.13^{*}$ \\
$\cdots$ & 6363.776 & 0.020 & $-10.303$ & $-2.17$ &$-7.83^{*}$ & $-8.13^{*}$ \\
10 & 6158.149 & 10.741 & $-1.891$ & 7.62 & $-3.96$ & $-7.23^{*}$ \\
10 & 6158.172 & 10.741 & $-1.031$ & 7.62 & $-3.96$ & $-7.23^{*}$ \\
10 & 6158.187 & 10.741 & $-0.441$ & 7.62 & $-3.96$ & $-7.23^{*}$ \\
\hline
\end{tabular}
\end{center}
Note ---
All data are were taken from Kurucz \& Bell's (1995) compilation
as far as available.\\
Followed by first four self-explanatory columns,
damping parameters are given in the last three columns:\\
Gammar is the radiation damping constant, $\log\gamma_{\rm rad}$.\\
Gammae is the Stark damping width per electron density
at $10^{4}$ K, $\log(\gamma_{\rm e}/N_{\rm e})$.\\
Gammaw is the van der Waals damping width per hydrogen density
at $10^{4}$ K, $\log(\gamma_{\rm w}/N_{\rm H})$. \\
$^{*}$ Computed as default values in the Kurucz's WIDTH program
(cf. Leusin, Topil'skaya 1987).
\end{table}


\setcounter{table}{2}
\begin{table}[h]
\caption{Coefficients ($a$, $b$) for O I 7771.94, 7774.17, and 7775.39.}
\tiny
\begin{center}
\begin{tabular}{cccr@{  }r@{ }cr@{  }r@{ }cr@{  }r@{ }c}\\
\hline\hline
Line & $T_{\rm eff}$ & $\log g$ &
$a$ ($\xi = 1$) & $b$ ($\xi = 1$) & $W$ range &
$a$ ($\xi = 2$) & $b$ ($\xi = 2$) & $W$ range &
$a$ ($\xi = 3$) & $b$ ($\xi = 3$) & $W$ range \\
\hline
7771.94 & 4500 & 1.0 & $-$0.075002 &  0.009542 &(~1,~65) & $-$0.075356 &  0.008023 &(~1,~74) & $-$0.074381 &  0.006728 &(~1,~85) \\[-1.0mm]
7771.94 & 4500 & 2.0 & $-$0.063934 &  0.009469 &(~0,~41) & $-$0.063630 &  0.008285 &(~0,~46) & $-$0.063137 &  0.006891 &(~0,~52) \\[-1.0mm]
7771.94 & 4500 & 3.0 & $-$0.035192 &  0.011033 &(~0,~27) & $-$0.034745 &  0.011525 &(~0,~28) & $-$0.034471 &  0.009788 &(~0,~30) \\[-1.0mm]
7771.94 & 4500 & 4.0 & $-$0.017092 &  0.007009 &(~0,~15) & $-$0.016955 &  0.007648 &(~0,~16) & $-$0.017007 &  0.006311 &(~0,~17) \\[-1.0mm]
7771.94 & 4500 & 5.0 & $-$0.007669 & $-$0.012856 &(~0,~~8) & $-$0.007191 & $-$0.014193 &(~0,~~8) & $-$0.006915 & $-$0.013005 &(~0,~~8) \\[-1.0mm]
7771.94 & 5000 & 1.0 & $-$0.107305 &  0.006369 &(~2,135) & $-$0.107536 &  0.005386 &(~2,155) & $-$0.106370 &  0.004483 &(~2,182) \\[-1.0mm]
7771.94 & 5000 & 2.0 & $-$0.093306 &  0.007027 &(~1,~96) & $-$0.093021 &  0.005981 &(~1,110) & $-$0.092000 &  0.005021 &(~1,126) \\[-1.0mm]
7771.94 & 5000 & 3.0 & $-$0.060149 &  0.008351 &(~0,~66) & $-$0.060438 &  0.007135 &(~0,~74) & $-$0.059794 &  0.006105 &(~0,~83) \\[-1.0mm]
7771.94 & 5000 & 4.0 & $-$0.026452 &  0.010508 &(~0,~44) & $-$0.026505 &  0.009854 &(~0,~48) & $-$0.026579 &  0.008254 &(~0,~52) \\[-1.0mm]
7771.94 & 5000 & 5.0 & $-$0.010772 &  0.007419 &(~0,~31) & $-$0.010895 &  0.006636 &(~0,~32) & $-$0.010664 &  0.006359 &(~0,~34) \\[-1.0mm]
7771.94 & 5500 & 1.0 & $-$0.173154 &  0.004333 &(~5,200) & $-$0.171278 &  0.003712 &(~5,234) & $-$0.168826 &  0.003108 &(~5,282) \\[-1.0mm]
7771.94 & 5500 & 2.0 & $-$0.130188 &  0.005075 &(~2,162) & $-$0.129085 &  0.004355 &(~2,186) & $-$0.127260 &  0.003656 &(~2,219) \\[-1.0mm]
7771.94 & 5500 & 3.0 & $-$0.088392 &  0.005886 &(~1,123) & $-$0.088652 &  0.005063 &(~1,138) & $-$0.087546 &  0.004300 &(~1,158) \\[-1.0mm]
7771.94 & 5500 & 4.0 & $-$0.045400 &  0.006999 &(~0,~89) & $-$0.045010 &  0.006242 &(~0,~98) & $-$0.044836 &  0.005401 &(~0,110) \\[-1.0mm]
7771.94 & 5500 & 5.0 & $-$0.017273 &  0.007421 &(~0,~66) & $-$0.017193 &  0.006644 &(~0,~71) & $-$0.017248 &  0.005989 &(~0,~76) \\[-1.0mm]
7771.94 & 6000 & 1.0 & $-$0.266939 &  0.003000 &(13,251) & $-$0.260351 &  0.002631 &(12,295) & $-$0.247962 &  0.002325 &(13,347) \\[-1.0mm]
7771.94 & 6000 & 2.0 & $-$0.199517 &  0.003669 &(~6,214) & $-$0.196006 &  0.003214 &(~6,245) & $-$0.191289 &  0.002757 &(~6,288) \\[-1.0mm]
7771.94 & 6000 & 3.0 & $-$0.129293 &  0.004429 &(~3,174) & $-$0.128434 &  0.003859 &(~2,200) & $-$0.126288 &  0.003282 &(~3,234) \\[-1.0mm]
7771.94 & 6000 & 4.0 & $-$0.069964 &  0.005137 &(~1,138) & $-$0.070334 &  0.004515 &(~1,155) & $-$0.068899 &  0.003922 &(~1,178) \\[-1.0mm]
7771.94 & 6000 & 5.0 & $-$0.029232 &  0.005318 &(~0,112) & $-$0.028856 &  0.004924 &(~0,120) & $-$0.029252 &  0.004288 &(~0,132) \\[-1.0mm]
7771.94 & 6500 & 1.0 & $-$0.425820 &  0.002043 &(25,282) & $-$0.406389 &  0.001874 &(25,331) & $-$0.382757 &  0.001703 &(25,389) \\[-1.0mm]
7771.94 & 6500 & 2.0 & $-$0.316286 &  0.002456 &(13,251) & $-$0.305817 &  0.002219 &(13,295) & $-$0.291792 &  0.001979 &(13,347) \\[-1.0mm]
7771.94 & 6500 & 3.0 & $-$0.198270 &  0.003252 &(~6,219) & $-$0.195040 &  0.002890 &(~6,251) & $-$0.187683 &  0.002557 &(~6,288) \\[-1.0mm]
7771.94 & 6500 & 4.0 & $-$0.106599 &  0.003996 &(~2,182) & $-$0.104490 &  0.003609 &(~2,204) & $-$0.102594 &  0.003146 &(~2,234) \\[-1.0mm]
7771.94 & 6500 & 5.0 & $-$0.044660 &  0.004404 &(~1,155) & $-$0.045000 &  0.003978 &(~1,170) & $-$0.044197 &  0.003590 &(~1,186) \\
\hline
7774.17 & 4500 & 1.0 & $-$0.073793 &  0.010148 &(~0,~58) & $-$0.073780 &  0.008488 &(~0,~66) & $-$0.072979 &  0.007092 &(~0,~76) \\[-1.0mm]
7774.17 & 4500 & 2.0 & $-$0.063712 &  0.006132 &(~0,~35) & $-$0.062821 &  0.008346 &(~0,~40) & $-$0.062356 &  0.006906 &(~0,~45) \\[-1.0mm]
7774.17 & 4500 & 3.0 & $-$0.035064 &  0.010989 &(~0,~23) & $-$0.034614 &  0.011527 &(~0,~23) & $-$0.034347 &  0.009430 &(~0,~26) \\[-1.0mm]
7774.17 & 4500 & 4.0 & $-$0.016938 &  0.007280 &(~0,~13) & $-$0.017033 &  0.005749 &(~0,~13) & $-$0.016773 &  0.004774 &(~0,~14) \\[-1.0mm]
7774.17 & 4500 & 5.0 & $-$0.007297 & $-$0.010223 &(~0,~~6) & $-$0.006829 & $-$0.012540 &(~0,~~6) & $-$0.007200 & $-$0.023415 &(~0,~~7) \\[-1.0mm]
7774.17 & 5000 & 1.0 & $-$0.105560 &  0.006706 &(~1,123) & $-$0.105520 &  0.005653 &(~1,141) & $-$0.104210 &  0.004709 &(~1,166) \\[-1.0mm]
7774.17 & 5000 & 2.0 & $-$0.091507 &  0.007408 &(~1,~87) & $-$0.091394 &  0.006292 &(~1,~98) & $-$0.090448 &  0.005256 &(~1,112) \\[-1.0mm]
7774.17 & 5000 & 3.0 & $-$0.059339 &  0.008759 &(~0,~58) & $-$0.059182 &  0.007595 &(~0,~65) & $-$0.058655 &  0.006327 &(~0,~72) \\[-1.0mm]
7774.17 & 5000 & 4.0 & $-$0.026360 &  0.010633 &(~0,~37) & $-$0.026406 &  0.009823 &(~0,~41) & $-$0.026522 &  0.008324 &(~0,~45) \\[-1.0mm]
7774.17 & 5000 & 5.0 & $-$0.010885 &  0.006427 &(~0,~26) & $-$0.010977 &  0.005904 &(~0,~26) & $-$0.010598 &  0.005227 &(~0,~28) \\[-1.0mm]
7774.17 & 5500 & 1.0 & $-$0.168793 &  0.004559 &(~4,191) & $-$0.166332 &  0.003922 &(~4,219) & $-$0.163428 &  0.003289 &(~4,257) \\[-1.0mm]
7774.17 & 5500 & 2.0 & $-$0.126021 &  0.005383 &(~2,148) & $-$0.125805 &  0.004568 &(~2,170) & $-$0.124542 &  0.003804 &(~2,200) \\[-1.0mm]
7774.17 & 5500 & 3.0 & $-$0.087354 &  0.006180 &(~1,110) & $-$0.086594 &  0.005325 &(~1,123) & $-$0.085991 &  0.004476 &(~1,141) \\[-1.0mm]
7774.17 & 5500 & 4.0 & $-$0.044505 &  0.007448 &(~0,~78) & $-$0.044722 &  0.006527 &(~0,~85) & $-$0.044613 &  0.005536 &(~0,~96) \\[-1.0mm]
7774.17 & 5500 & 5.0 & $-$0.016934 &  0.007810 &(~0,~56) & $-$0.017195 &  0.006915 &(~0,~60) & $-$0.017136 &  0.006154 &(~0,~65) \\[-1.0mm]
7774.17 & 6000 & 1.0 & $-$0.254642 &  0.003237 &(~9,240) & $-$0.243609 &  0.002897 &(~9,275) & $-$0.236696 &  0.002478 &(~9,331) \\[-1.0mm]
7774.17 & 6000 & 2.0 & $-$0.192306 &  0.003910 &(~4,200) & $-$0.189738 &  0.003393 &(~4,229) & $-$0.185979 &  0.002875 &(~4,269) \\[-1.0mm]
7774.17 & 6000 & 3.0 & $-$0.125586 &  0.004661 &(~2,162) & $-$0.124640 &  0.004046 &(~2,182) & $-$0.122884 &  0.003402 &(~2,214) \\[-1.0mm]
7774.17 & 6000 & 4.0 & $-$0.068886 &  0.005361 &(~1,126) & $-$0.068406 &  0.004741 &(~1,138) & $-$0.067629 &  0.004055 &(~1,155) \\[-1.0mm]
7774.17 & 6000 & 5.0 & $-$0.028857 &  0.005626 &(~0,~96) & $-$0.028811 &  0.005050 &(~0,105) & $-$0.028577 &  0.004510 &(~0,112) \\[-1.0mm]
7774.17 & 6500 & 1.0 & $-$0.390958 &  0.002323 &(19,269) & $-$0.375496 &  0.002093 &(18,316) & $-$0.356457 &  0.001876 &(19,372) \\[-1.0mm]
7774.17 & 6500 & 2.0 & $-$0.294824 &  0.002727 &(~9,240) & $-$0.285213 &  0.002443 &(~9,275) & $-$0.272451 &  0.002166 &(~9,324) \\[-1.0mm]
7774.17 & 6500 & 3.0 & $-$0.189861 &  0.003492 &(~4,204) & $-$0.185155 &  0.003121 &(~4,229) & $-$0.181040 &  0.002675 &(~4,269) \\[-1.0mm]
7774.17 & 6500 & 4.0 & $-$0.101634 &  0.004302 &(~2,166) & $-$0.101045 &  0.003789 &(~2,186) & $-$0.099058 &  0.003275 &(~2,214) \\[-1.0mm]
7774.17 & 6500 & 5.0 & $-$0.043878 &  0.004657 &(~1,138) & $-$0.043743 &  0.004213 &(~1,148) & $-$0.043188 &  0.003693 &(~1,166) \\
\hline
7775.39 & 4500 & 1.0 & $-$0.071843 &  0.010860 &(~0,~48) & $-$0.071891 &  0.009057 &(~0,~55) & $-$0.070976 &  0.007520 &(~0,~62) \\[-1.0mm]
7775.39 & 4500 & 2.0 & $-$0.061323 &  0.009180 &(~0,~29) & $-$0.060937 &  0.008329 &(~0,~32) & $-$0.060347 &  0.006721 &(~0,~36) \\[-1.0mm]
7775.39 & 4500 & 3.0 & $-$0.034389 &  0.010752 &(~0,~18) & $-$0.033807 &  0.010977 &(~0,~18) & $-$0.033636 &  0.008536 &(~0,~20) \\[-1.0mm]
7775.39 & 4500 & 4.0 & $-$0.016737 &  0.002727 &(~0,~10) & $-$0.016927 &  0.000833 &(~0,~10) & $-$0.016618 &  0.001652 &(~0,~10) \\[-1.0mm]
7775.39 & 4500 & 5.0 & $-$0.010138 & $-$0.096676 &(~0,~~4) & $-$0.010362 & $-$0.095794 &(~0,~~4) & $-$0.009923 & $-$0.090568 &(~0,~~4) \\[-1.0mm]
7775.39 & 5000 & 1.0 & $-$0.102353 &  0.007223 &(~1,107) & $-$0.102838 &  0.006040 &(~1,123) & $-$0.101612 &  0.005036 &(~1,141) \\[-1.0mm]
7775.39 & 5000 & 2.0 & $-$0.089035 &  0.007937 &(~0,~72) & $-$0.089088 &  0.006640 &(~0,~83) & $-$0.088580 &  0.005489 &(~0,~93) \\[-1.0mm]
7775.39 & 5000 & 3.0 & $-$0.058042 &  0.008875 &(~0,~47) & $-$0.058165 &  0.007677 &(~0,~52) & $-$0.057643 &  0.006315 &(~0,~59) \\[-1.0mm]
7775.39 & 5000 & 4.0 & $-$0.025786 &  0.011203 &(~0,~30) & $-$0.026091 &  0.009942 &(~0,~32) & $-$0.025790 &  0.008631 &(~0,~35) \\[-1.0mm]
7775.39 & 5000 & 5.0 & $-$0.010567 &  0.006446 &(~0,~19) & $-$0.010740 &  0.004815 &(~0,~20) & $-$0.010665 &  0.003706 &(~0,~20) \\[-1.0mm]
7775.39 & 5500 & 1.0 & $-$0.161349 &  0.004926 &(~2,170) & $-$0.160227 &  0.004180 &(~2,195) & $-$0.158235 &  0.003473 &(~2,229) \\[-1.0mm]
7775.39 & 5500 & 2.0 & $-$0.121987 &  0.005726 &(~1,129) & $-$0.122434 &  0.004800 &(~1,148) & $-$0.120899 &  0.003995 &(~1,170) \\[-1.0mm]
7775.39 & 5500 & 3.0 & $-$0.084362 &  0.006543 &(~0,~93) & $-$0.084568 &  0.005556 &(~0,105) & $-$0.083895 &  0.004600 &(~0,117) \\[-1.0mm]
7775.39 & 5500 & 4.0 & $-$0.043955 &  0.007729 &(~0,~63) & $-$0.044054 &  0.006737 &(~0,~69) & $-$0.043452 &  0.005682 &(~0,~76) \\[-1.0mm]
7775.39 & 5500 & 5.0 & $-$0.017604 & $-$0.001526 &(~0,~44) & $-$0.017034 &  0.006931 &(~0,~47) & $-$0.017098 &  0.006145 &(~0,~49) \\[-1.0mm]
7775.39 & 6000 & 1.0 & $-$0.230850 &  0.003678 &(~6,219) & $-$0.226046 &  0.003208 &(~6,251) & $-$0.220972 &  0.002705 &(~6,302) \\[-1.0mm]
7775.39 & 6000 & 2.0 & $-$0.182907 &  0.004231 &(~3,178) & $-$0.182280 &  0.003577 &(~3,209) & $-$0.179144 &  0.003005 &(~3,240) \\[-1.0mm]
7775.39 & 6000 & 3.0 & $-$0.120395 &  0.004954 &(~1,141) & $-$0.120003 &  0.004222 &(~1,158) & $-$0.118602 &  0.003513 &(~1,182) \\[-1.0mm]
7775.39 & 6000 & 4.0 & $-$0.066570 &  0.005688 &(~0,105) & $-$0.066206 &  0.004914 &(~0,115) & $-$0.065479 &  0.004137 &(~0,129) \\[-1.0mm]
7775.39 & 6000 & 5.0 & $-$0.027805 &  0.006041 &(~0,~78) & $-$0.027918 &  0.005307 &(~0,~83) & $-$0.028125 &  0.004557 &(~0,~91) \\[-1.0mm]
7775.39 & 6500 & 1.0 & $-$0.348779 &  0.002725 &(12,251) & $-$0.337745 &  0.002434 &(12,288) & $-$0.328159 &  0.002095 &(12,347) \\[-1.0mm]
7775.39 & 6500 & 2.0 & $-$0.266789 &  0.003117 &(~6,219) & $-$0.260165 &  0.002749 &(~6,251) & $-$0.252698 &  0.002368 &(~6,295) \\[-1.0mm]
7775.39 & 6500 & 3.0 & $-$0.178059 &  0.003805 &(~3,182) & $-$0.175237 &  0.003326 &(~2,204) & $-$0.172827 &  0.002783 &(~2,240) \\[-1.0mm]
7775.39 & 6500 & 4.0 & $-$0.096075 &  0.004634 &(~1,141) & $-$0.096248 &  0.003982 &(~1,158) & $-$0.094877 &  0.003362 &(~1,182) \\[-1.0mm]
7775.39 & 6500 & 5.0 & $-$0.042435 &  0.004943 &(~0,112) & $-$0.042258 &  0.004382 &(~0,123) & $-$0.041729 &  0.003805 &(~0,135) \\
\hline
\end{tabular}
\end{center}
Note --- The coefficients ($a$, $b$) used with the analytic formula
 [Eq.(1)] for evaluating the non-LTE correction, corresponding to each of 
the atmospheric models ($T_{\rm eff}$ and $\log g$ values shown in columns
2 and 3) for three values of the microturbulent velocity ($\xi$ = 1, 2, and 
3 km~s$^{-1}$).
The ``$W$ range'' gives the minimum and the maximum value of 
the equivalent-width data (in m$\rm\AA$) based on which the coefficients
were established (i.e., indication of the range of the $W_{\lambda}$ values
to which the formula with these coefficients can be applied).
\end{table}

\clearpage
\small
\setcounter{table}{3}
\begin{table}[h]
\caption{Non-LTE corrections of O~{\sc i} 7771.94 line for the same 
equivalent width but with different metallicities.}
\begin{center}
\begin{tabular}
{ccrrcrrrrrrc}\hline \hline
$T_{\rm eff}$ & $\log g$ & $[X]$ & [O/Fe] & $\xi$ & $W_{7771}$ &
$r_{0}^{\rm NLTE}$  & $\overline{\log \tau}$ & $\log \epsilon^{\rm NLTE}$ & 
$\log \epsilon^{\rm LTE}$ & $\Delta\log\epsilon$ & 
$\Delta\log\epsilon_{\rm Gratton}$ \\
(K) & (cm~s$^{-2}$) &  &  & (km~s$^{-1}$) & (m$\rm\AA$) & & & & & & \\
\hline
6000 & 3.0 & 0 & $-0.13$ & 2. & 152. & 0.378 & $-1.05$ & 8.80 & 9.30 & $-0.50$ & $-0.13$ \\
6000 & 3.0 & $-1$ & $+0.86$ & 2. & 152. & 0.383 & $-1.13$ & 8.79 & 9.35 & $-0.56$ & $-0.45$ \\
\hline
6000 & 3.0 & $-2$ & $+0.21$ & 2. & 20. & 0.877 & $-0.40$ & 7.14 & 7.29 & $-0.15$ & $-0.14$ \\
6000 & 3.0 & $-3$ & $+1.23$ & 2. & 20. & 0.878 & $-0.40$ & 7.16 & 7.31 & $-0.15$ & $-0.30$ \\
\hline
\end{tabular}
\end{center}
Note --- Shown are the NLTE corrections for the $T_{\rm eff}$ = 6000~K 
and  $\log g = 3.0$ model atmospheres of four different metallicities 
([$X$] = [Fe/H]) of 0, $-1$, $-2$, and $-3$, but the oxygen abundances 
were carefully adjusted to yield the same O~{\sc i} 7771.94 equivalent 
widths $W$ = 152 m$\rm\AA$ (for [$X$] = 0 and $-1$) and 
$W$ = 20 m$\rm\AA$ (for [$X$] = $-2$ and $-3$) in order to enable
the direct comparison with Gratton et al.'s (1999) results.
$r_{0}^{\rm NLTE}$ is the residual flux at the line center 
and $\overline{\log \tau}$ is the mean-depth of line formation 
[defined as $\int R_{\lambda} \log\tau_{5000}(\tau_{\lambda}=2/3) d\lambda / 
\int R_{\lambda} d\lambda$,
where $R_{\lambda}$ ($\equiv 1 - r_{\lambda}$) is the line-depth]; 
other symbols are self-explanatory.
Note that our NLTE corrections ($\Delta\log\epsilon$; column 11) are 
practically the same for the [$X$] = 0 and $-1$ cases as well as 
for the [$X$] = $-2$ and $-3$ cases, in marked contrast with Gratton et al.'s
(1999) results ($\Delta\log\epsilon_{\rm Gratton}$; shown in column 12 
for a comparison).
\end{table}


\clearpage
\setcounter{table}{5}
\begin{table}[h]
\small
\caption{[O/Fe] ratio averaged for each metallicity bin.}
\begin{center}
\begin{tabular}
{cc@{  }c@{  }cc@{  }c@{  }cc}\hline\hline
Metallicity Range  & $n_{77}$ & $\langle[$O/Fe$]_{77}\rangle$ & $\sigma_{77}$
                   & $n_{63}$ & $\langle[$O/Fe$]_{63}\rangle$ & $\sigma_{63}$
                   & $\langle[$O/Fe$]_{77}\rangle - \langle[$O/Fe$]_{63}\rangle$\\
\hline
$-2.75 \leq$ [Fe/H] $< -2.25$ & 31 & +0.83 & 0.22 & 12 & +0.59 & 0.15 & +0.24\\
$-2.25 \leq$ [Fe/H] $< -1.75$ & 31 & +0.75 & 0.23 & 19 & +0.45 & 0.17 & +0.30\\
$-1.75 \leq$ [Fe/H] $< -1.25$ & 37 & +0.69 & 0.24 & 32 & +0.46 & 0.17 & +0.25\\
$-1.25 \leq$ [Fe/H] $< -0.75$ & 34 & +0.54 & 0.23 & 19 & +0.36 & 0.21 & +0.18\\
$-0.75 \leq$ [Fe/H] $< -0.25$ & 54 & +0.26 & 0.16 & 20 & +0.18 & 0.22 & +0.08\\
$-0.25 \leq$ [Fe/H] $< +0.25$ & 28 & $-$0.04 & 0.12 & 10 & +0.14 & 0.16 & $-$0.18\\
\hline
\end{tabular}
\end{center} 
Note ---
$n_{77}$ and $n_{63}$ are the numbers of stars used for calculating
$\langle[$O/Fe$]_{77}\rangle$ and $\langle[$O/Fe$]_{63}\rangle$ at 
each metallicity bin, while $\sigma_{77}$ and $\sigma_{63}$ are the 
standard deviations.\\
\end{table}

\clearpage

\onecolumn

\begin{figure}
\centering
\includegraphics[width=0.9\textwidth]{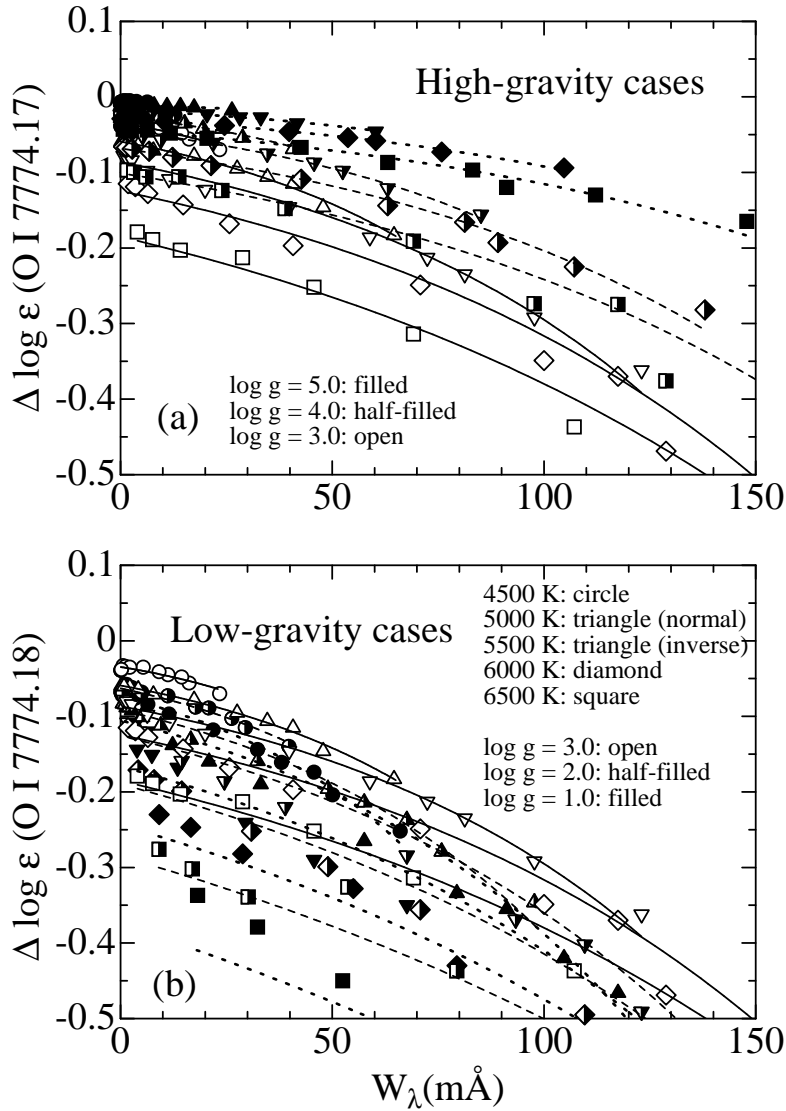}
\caption{
Non-LTE correction vs. equivalent-width relation
calculated for O~{\sc i} 7774.17 (the middle line of the oxygen triplet)
with the microturbulent velocity ($\xi$) of 2 \kms , as constructed from
($W_{i}^{\rm NLTE}$, $\Delta_{i}$) given in Table 2. Results corresponding 
to each of the models are discerned by differences in symbols.
(a): high gravity cases ($\log g$ = 5.0, 4.0, and 3.0); 
(b): low gravity cases ($\log g$ = 3.0, 2.0, and 1.0).
The curves drawn through the points show the analytical 
approximations [based on Eq. (1) with the coefficients
given in Table 3], in which each line-type corresponds to the different
surface gravity case: $\log g$ = 5 (dotted), 4 (dashed), 3 (solid), 
2 (dashed), and 1 (dotted).
}
\label{Fig1}
\end{figure}

\begin{figure}
\centering
\includegraphics[width=0.9\textwidth]{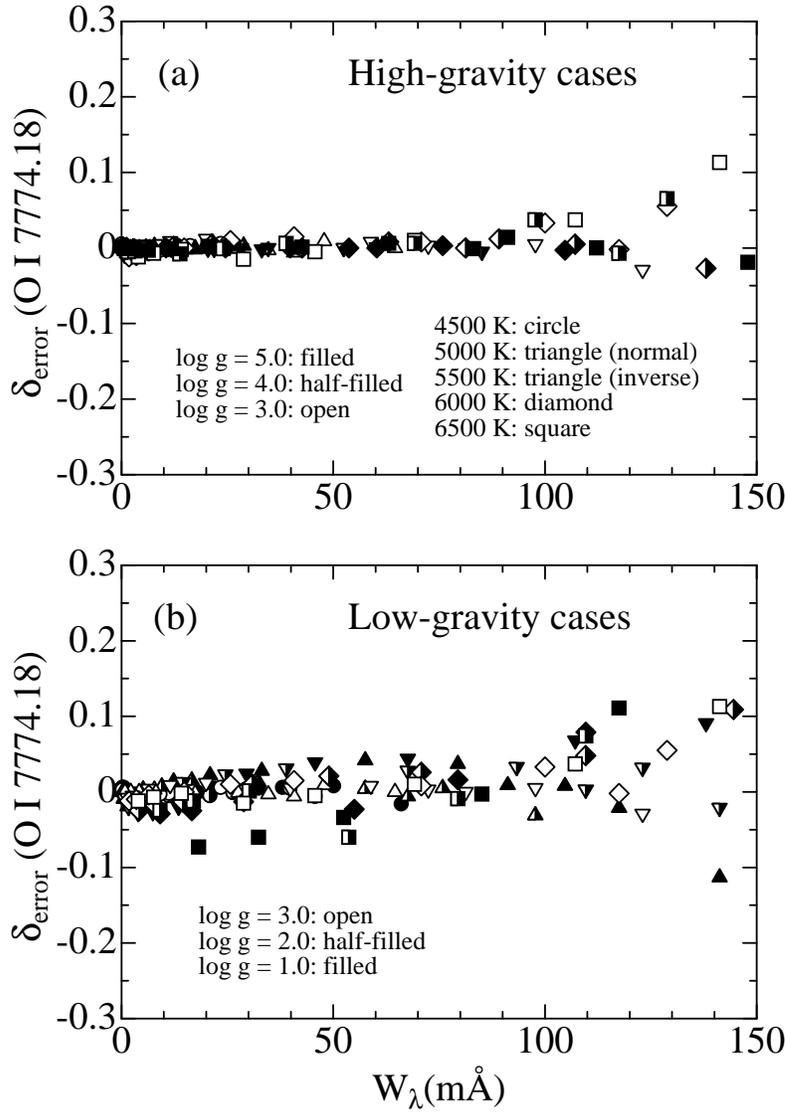}
\caption{Error of the analytical expression for the non-LTE
correction (O~{\sc i} 7774.17, $\xi$ = 2 \kms), 
$\Delta = a 10^{b W_{\lambda}}$ 
with the coefficients ($a, b$) given in Table 3, 
relative to the exactly calculated value (Table 2). 
Similarly to Fig. 1, the results are shown as functions of $W_{\lambda}$.
}
\label{Fig2}
\end{figure}

\begin{figure}
\centering
\includegraphics[width=0.9\textwidth]{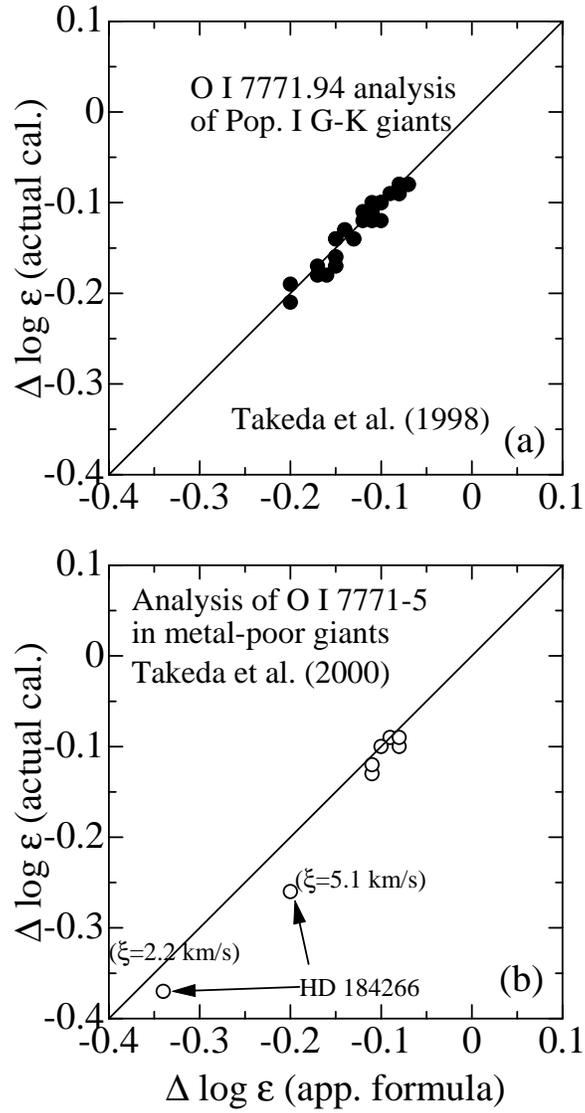}
\caption{Comparison of the non-LTE abundance corrections
derived in previous two works of the author, with those calculated
by our analytical formula [Eq. (1)] using appropriate ($a, b$) coefficients 
evaluated by interpolating the values in Table 3.
(a): O~{\sc i} 7771.94 in population I G--K giants (Takeda et al. 1998);
(b): O~{\sc i} 7771--5 in very metal-poor giants (Takeda et al. 2000).
}
\label{Fig3}
\end{figure}

\begin{figure}
\centering
\includegraphics[width=0.9\textwidth]{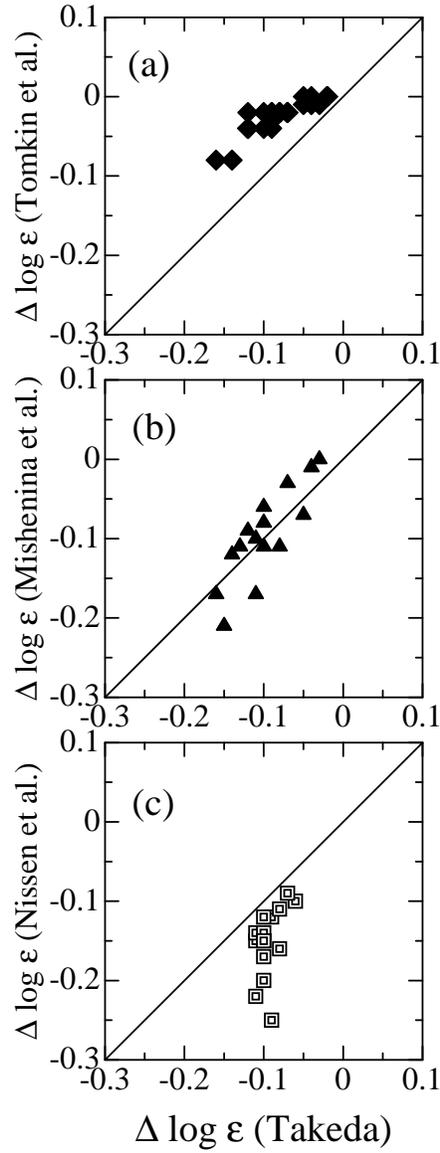}
\caption{Comparison of our non-LTE abundance corrections
for O~{\sc i} 7771--5 triplet resulting from reanalysis of the published
equivalent-width data by use of our approximate analytical formula 
with the original values determined by those authors.
(a): Tomkin et al. (1992); (b): Mishenina et al. (2000);
(c): Nissen et al. (2002).
}
\label{Fig4}
\end{figure}

\begin{figure}
\centering
\includegraphics[width=0.9\textwidth]{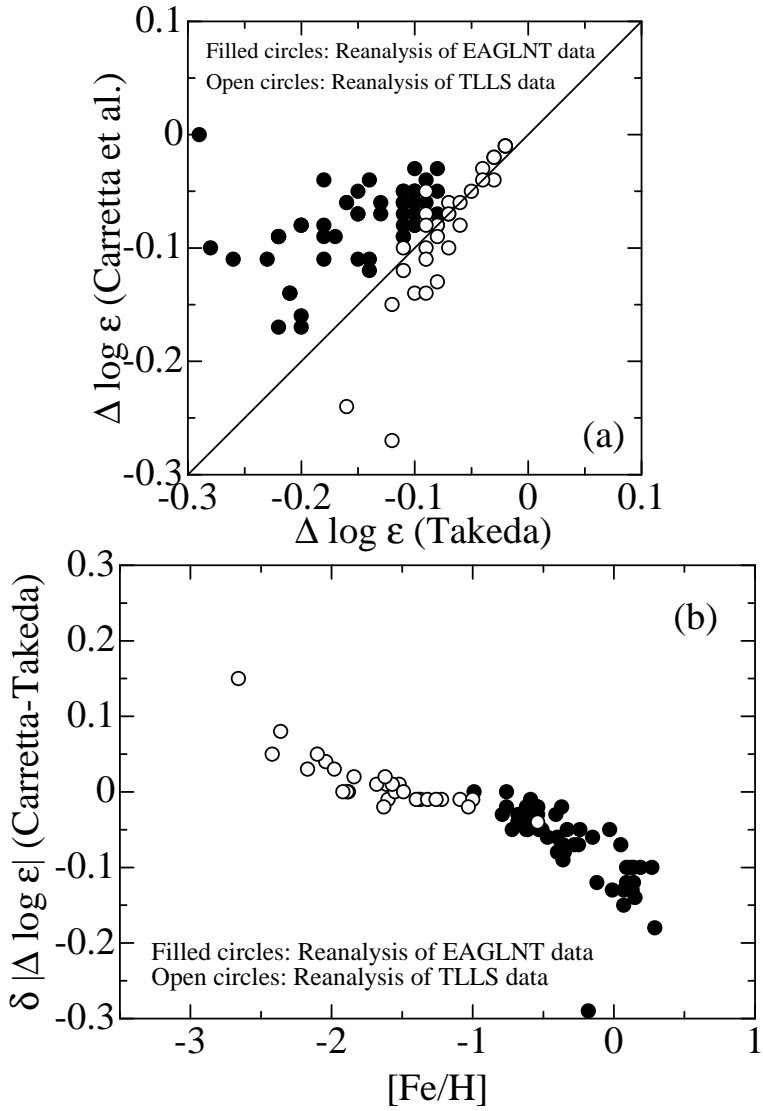}
\caption{(a): Comparison of our non-LTE abundance corrections
for O~{\sc i} 7771--5 triplet resulting from reanalysis of the data of 
Tomkin et al. (1992; open circles) and Edvardsson et al. 
(1993; closed circles) (as was done by Carretta et al. 2000) 
with the corrections determined by Carretta et al. (2000). 
(b): Differences between Carretta et al.'s (2000)
non-LTE corrections and ours, plotted as functions of metallicity.
}
\label{Fig5}
\end{figure}

\begin{figure}
\centering
\includegraphics[width=0.9\textwidth]{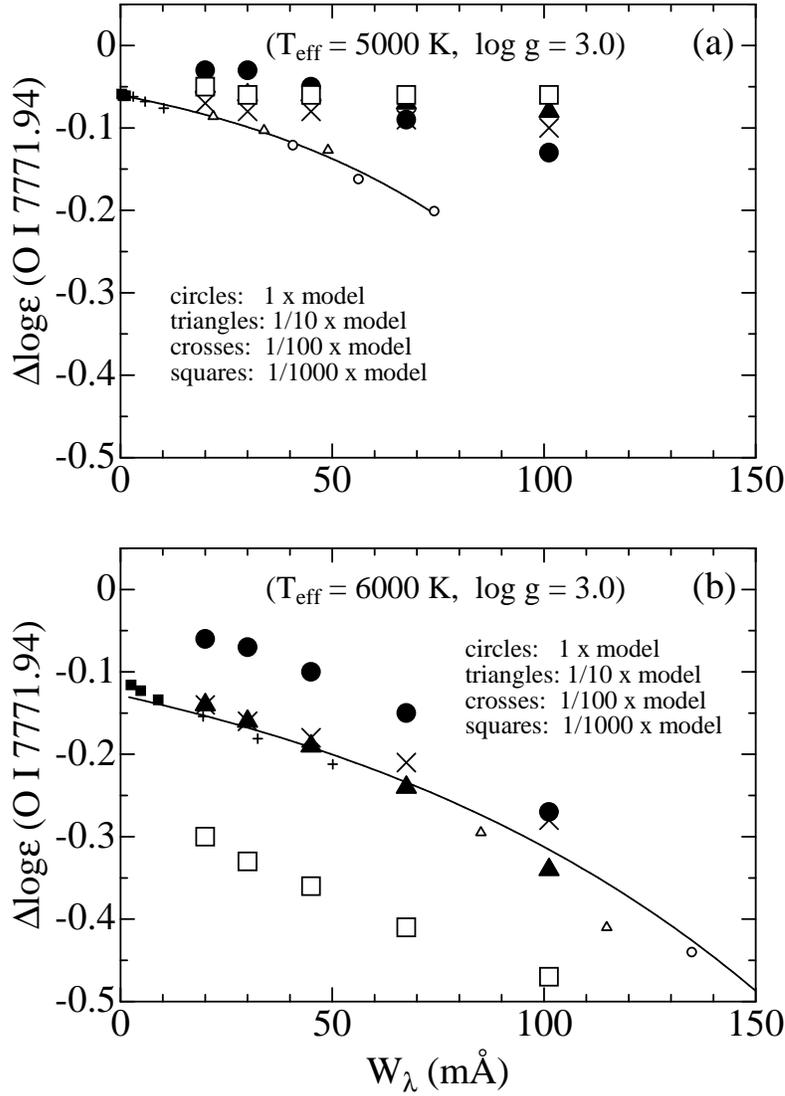}
\caption{
Non-LTE correction vs. equivalent-width relations for 
the O~{\sc i} 7771.94 line. 
Large symbols show the results calculated by Gratton et al. (1999), 
while small symbols indicate our results taken from Table 2.
The model metallicity corresponding to each data point is denoted by 
the symbol type as explained in the figure.
Our relevant (metallicity-independent) analytical relation [Eq. (1)
coupled with Table 3] is depicted by the solid line, which was 
determined so as to fit the original data (small symbols).
(a): $T_{\rm ef}$ = 5000 K, $\log g$ = 3.0;
(b): $T_{\rm ef}$ = 6000 K, $\log g$ = 3.0.
}
\label{Fig6}
\end{figure}

\begin{figure}
\centering
\includegraphics[width=0.9\textwidth]{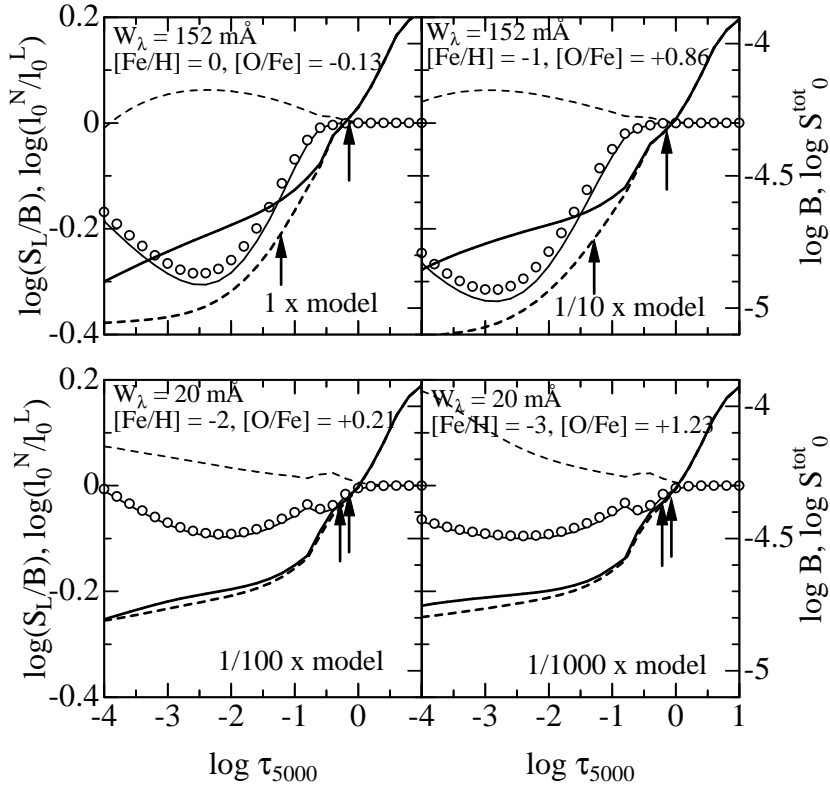}
\caption{
Run of the physical quantities involved with the line formation
of the O~{\sc i} 7771.94 line, for the four demonstrative cases
of the metallicity-dependence test given in Table 4.
The thin solid line and the thin dashed line indicate
the depth-dependence of the line source function (in unit of 
the local Planck function) and the NLTE-to-LTE line-opacity, 
respectively (corresponding to the scales marked in the left-hand axis). 
The corresponding pure two-level-atom solution of $S_{\rm L}/B$ 
[derived by solving the scattering problem with the line source function
of Eq. (2) while using the actual depth-dependent
$\epsilon'(\tau)$,  $\eta(\tau)$, and $B(\tau)$] is also shown
by the open circles, which is in good agreement with the full
non-LTE solution presented by Eq. (3) (thin solid line).
Meanwhile, the thick solid line and the thick dashed line show
the run of the Planck function itself and that of the total 
(i.e., line plus continuum) source function at the line center,
respectively (corresponding to the scales marked in the right-hand axis).
The depth points corresponding to the continuum optical-depth of unity
and the line-center optical depth of unity are indicated by two 
upward arrows.
}
\label{Fig7}
\end{figure}

\begin{figure}
\centering
\includegraphics[width=0.9\textwidth]{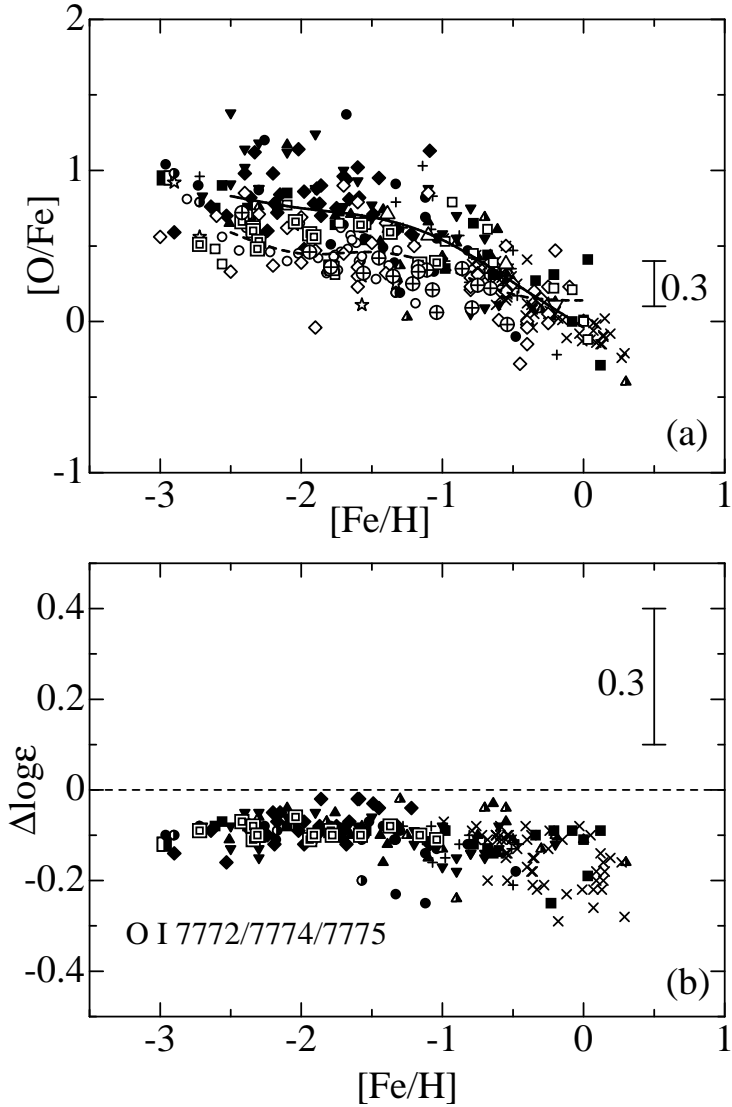}
\caption{(a): [O/Fe] vs. [Fe/H] relation based on the 
reanalysis of published equivalent width data of O~{\sc i} 7771--5 lines 
(closed symbols, half-filled symbols, crosses) and [O~{\sc i}] 6300/6363 lines 
(open symbols), where the non-LTE correction was taken into account only 
for the O~{\sc i} triplet lines.
The solid and dashed lines indicate the mean [O/Fe] tendencies 
suggested from permitted O~{\sc i} lines and forbidden [O~{\sc i}] lines, 
respectively. (b): Applied non-LTE corrections for the O~{\sc i}~7771--5 lines
plotted with respect to the metallicity.
Each of the symbols denote the sources of equivalent widths.
O~{\sc i} 7771--5: 
filled squares---Carretta et al. (2000), 
St. Andrew's crosses---Edvardsson et al. (1993),
filled inverse triangles---Abia \& Rebolo (1989),
Greek Crosses---Boesgaard \& King (1993),
half-filled triangles---Sneden et al. (1979),
filled triangles---Mishenina et al. (2000),
filled diamonds---Tomkin et al. (1992),
filled circles---Cavallo et al. (1997),
half-filled circles---Takeda et al. (2000),
half-filled square---King (1994),
double squares---Nissen et al. (2002).
[O~{\sc i}] 6300/6363:
open squares---Carretta et al. (2000),
open inverse triangles---Sneden et al. (1979),
open triangles---Mishenina et al. (2000),
open diamonds---Barbuy (1988) and Barbuy \& Erdelyi-Mendes (1989),
open circles---Sneden et al. (1991) and Kraft et al. (1992),
open stars---Takeda et al. (2000),
circled pluses---Nissen et al. (2002).
The original data (equivalent widths and atmospheric parameters) 
adopted from the literature and the results of the analysis 
(abundances and non-LTE corrections) are given in Table 5 
(available only electronically).}
\label{Fig8}
\end{figure}

\begin{figure}
\centering
\includegraphics[width=0.9\textwidth]{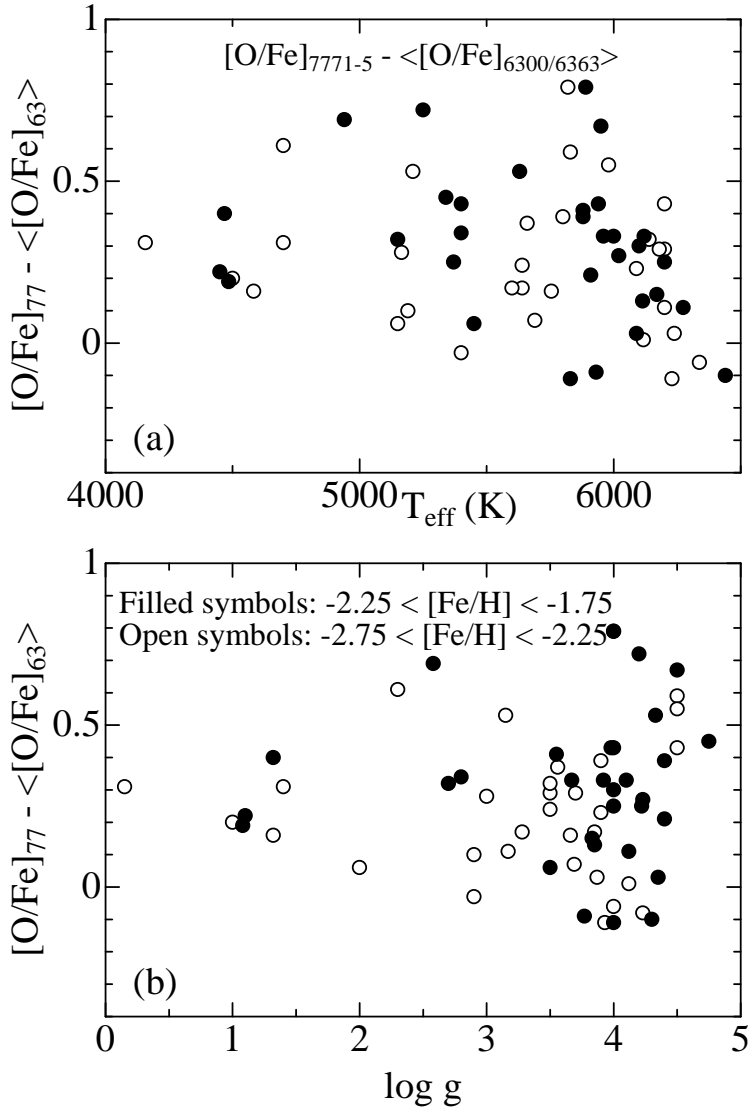}
\caption{Dependence of 
[O/Fe]$_{7771-5} - \langle[$O/Fe$]_{6300/6363}\rangle$ 
(the discrepancy of each star's [O/Fe]$_{7771-5}$ relative to the mean 
$\langle[$O/Fe$]_{6300/6363}\rangle$) upon (a) $T_{\rm eff}$ and (b) $\log g$.
Open symbols are for stars of $-2.75 \leq$ [Fe/H] $< -2.25$, while
filled symbols correspond to those of $-2.25 \leq$ [Fe/H] $< -1.75$.
}
\label{Fig9}
\end{figure}

\begin{figure}
\centering
\includegraphics[width=0.9\textwidth]{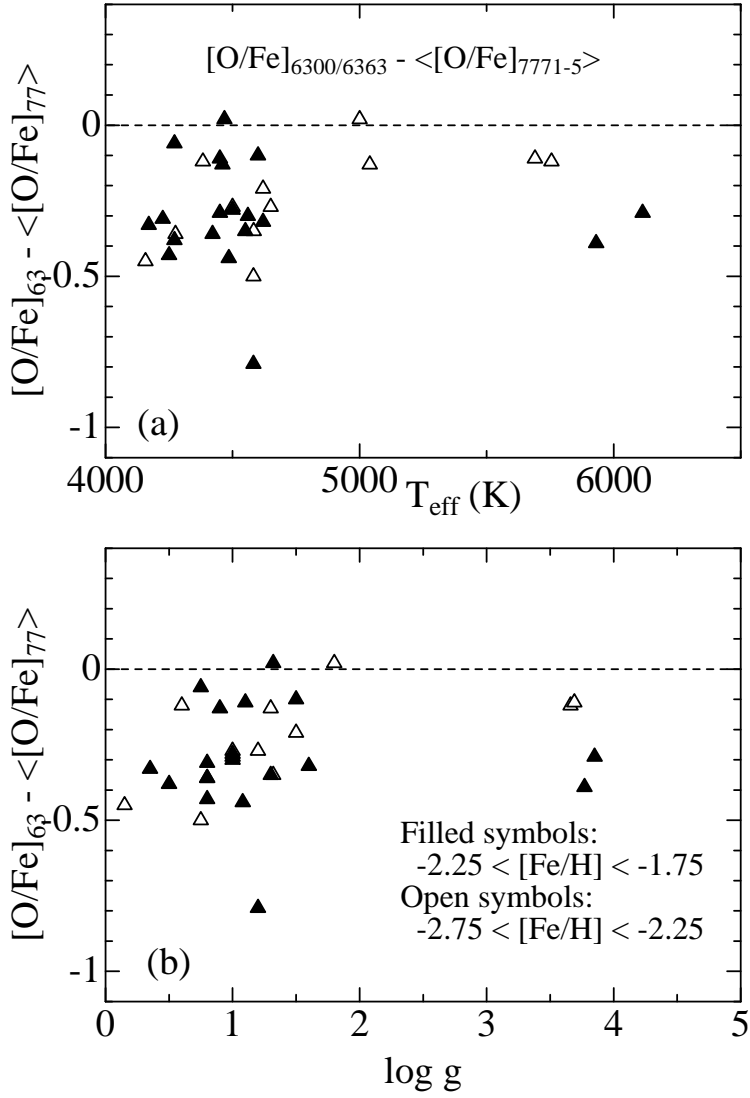}
\caption{Dependence of 
[O/Fe]$_{6300/6363} - \langle[$O/Fe$]_{7771-5}\rangle$
upon (a) $T_{\rm eff}$ and (b) $\log g$. 
Otherwise, the same as in Fig. 9.
}
\label{Fig10}
\end{figure}

\begin{figure}
\centering
\includegraphics[width=0.9\textwidth]{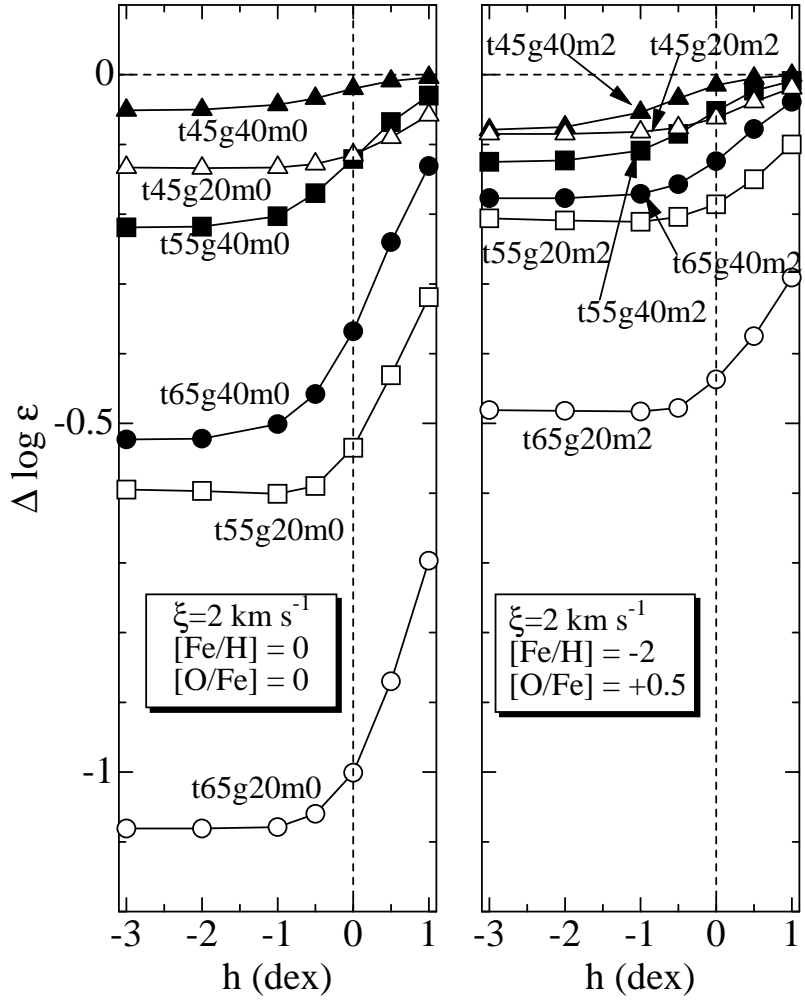}
\caption{Dependence of the non-LTE correction for
the O~{\sc i}~7774.18 line upon the logarithmic correction ($h$)
applied to the classical treatment of H~{\sc i} collision rates.
Left panel: ([O/Fe] = 0.0, $\xi$ = 2 km~s$^{-1}$) calculations for 
representative [Fe/H] = 0 models;
right panel: ([O/Fe] = +0.5, $\xi$ = 2 km~s$^{-1}$) calculations for 
representative [Fe/H] = $-2$ models.
Code ``t$aa$g$bb$m$c$'' denotes the model with 
$T_{\rm eff} ({\rm K}) = aa \times 100$,
$\log g$ (cm~s$^{-2}$) = $bb / 10$, and [Fe/H] (metallicity) = $- c$.
}
\label{Fig11}
\end{figure}

\begin{figure}
\centering
\includegraphics[width=0.9\textwidth]{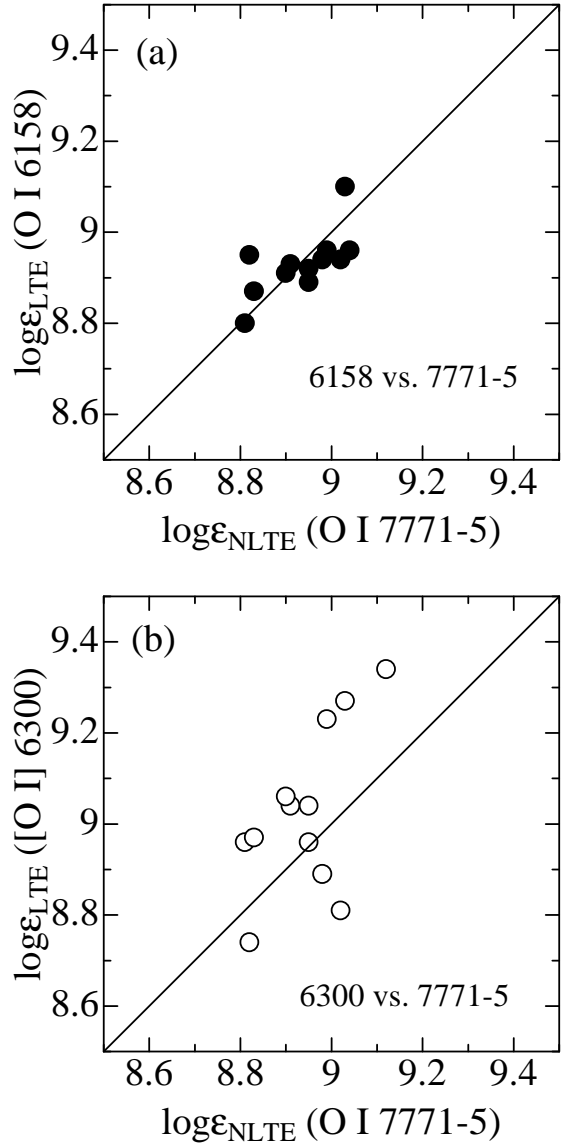}
\caption{Comparison of the oxygen abundances
derived from the O~{\sc i} 7771--5 lines (with the NLTE corrections
presented in this study) with those determined from other oxygen lines
for nearby solar-type stars analyzed by Takeda et al. (2001):
(a) O~{\sc i} 6158.2 line (comprising three components),
(b) [O~{\sc i}] 6300 forbidden line.
}
\label{Fig12}
\end{figure}

\begin{figure}
\centering
\includegraphics[width=0.9\textwidth]{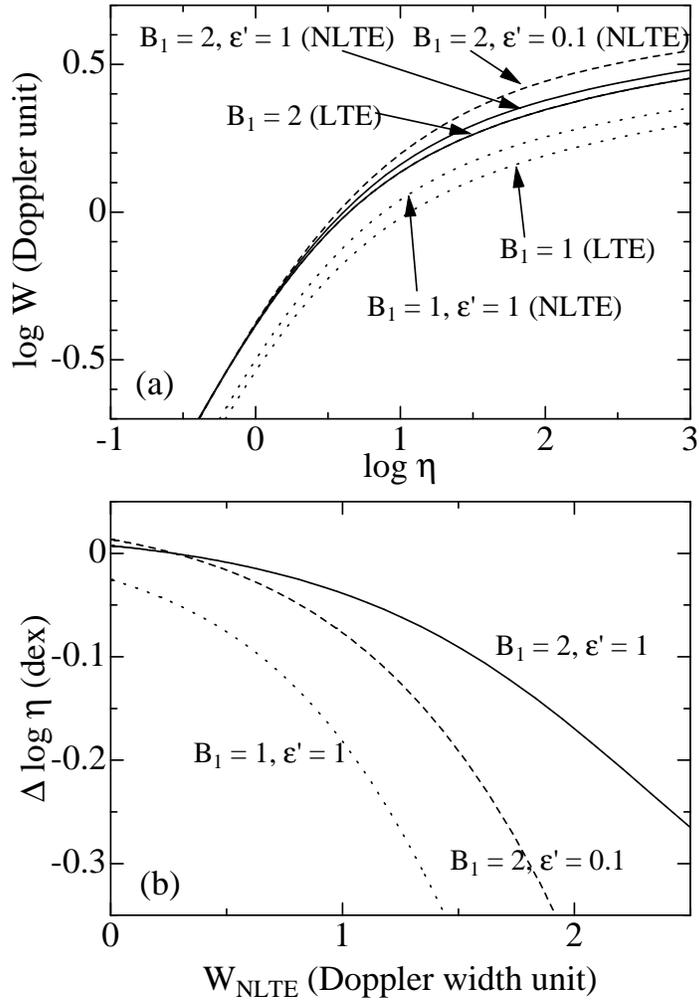}
\caption{Results of the calculations on the simple
``two-level atom'' model for simulating the
formation of the O~{\sc i} 7771--5 triplet. Shown are the results
computed for three different combinations of $B_{1}$ (gradient
of the Planck function) and $\epsilon'$ (photon destruction
probability) as indicated in the figure. (a) Theoretical curves of 
growth computed for LTE (without scattering) and NLTE
(with scattering). (b) Run of the non-LTE correction 
($\equiv \log\eta_{\rm NLTE} - \log\eta_{\rm LTE}$) with the NLTE
equivalent width.
}
\label{Fig13}
\end{figure}

\end{document}